\documentclass[traditabstract]{aa}
\usepackage{graphicx}
\usepackage{txfonts}
\usepackage{booktabs}
\usepackage{natbib}
\usepackage[utf8]{inputenc}
\bibpunct{(}{)}{;}{a}{}{,} 
\usepackage[usenames]{color}

\newcommand{\gcn}{GCN Circ.}

\newcommand{\swift}{\textit{Swift}}
\newcommand{\fermi}{\textit{Fermi}}

\usepackage{colortbl, xcolor}

\begin{document}

\title{The \swift/\fermi\ 
GRB 080928 from 1 eV to 150 keV}

\author{
A.~Rossi\inst{1},
S.~Schulze\inst{1,2},
S.~Klose\inst{1},
D.~A.~Kann\inst{1},
A.~Rau\inst{3},
H.~A.~Krimm\inst{4},
G.~J\'ohannesson\inst{5},
A.~Panaitescu\inst{6},
F.~Yuan\inst{7},
P.~Ferrero\inst{1,8,9},
T.~Kr\"uhler\inst{3,10},
J.~Greiner\inst{3},
P.~Schady\inst{3},
S.~B.~Pandey\inst{7,11},
L.~Amati\inst{12},
P.~M.~J.~Afonso\inst{3}\fnmsep\thanks{\emph{Present address: }
American River College, Physics Department, 4700 College Oak Drive,
Sacramento, CA 95841, USA},
C.~W.~Akerlof\inst{7},
L.~A.~Arnold\inst{13},
C.~Clemens\inst{3},
R.~Filgas\inst{3},
D.~H.~Hartmann\inst{14},
A.~K\"upc\"u~Yolda\c{s}\inst{3,15},
S.~McBreen\inst{3,16},
T.~A.~McKay\inst{7},
A.~Nicuesa~Guelbenzu\inst{1},
F.~Olivares~E.\inst{3},
B.~Paciesas\inst{17},
E.~S.~Rykoff\inst{18},
G.~Szokoly\inst{3,19},
A.~C.~Updike\inst{14,20,21},
\and
A.~Yolda\c{s}\inst{15}
}

\offprints{A. Rossi, rossi@tls-tautenburg.de}

\institute{Th\"uringer Landessternwarte Tautenburg, Sternwarte 5, D--07778 Tautenburg, Germany 
\and
   Centre for Astrophysics and Cosmology, Science Institute, University of Iceland, Dunhagi 5, IS-107 Reykjav\'ik, Iceland 
\and
   Max-Planck-Institut f\"ur Extraterrestrische Physik, Giessenbachstra\ss e, D--85748 Garching, Germany 
\and   
   CRESST, Universities Space Research Association and NASA GSFC, Greenbelt, MD 20771, USA 
\and
	 Hansen Experimental Physics Laboratory, Stanford University, Stanford, CA 94305, USA 
\and 
   ISR-1, Los Alamos National Laboratory, Los Alamos, NM 87545, USA 
\and 
   Physics Department, University of Michigan, Ann Arbor, MI 48109, USA 
\and
   Instituto de Astrof\'{\i}sica de Canarias (IAC), E-38200 La Laguna, Tenerife, Spain 
\and
   Departamento de Astrof\'{\i}sica, Universidad de La Laguna (ULL), E-38205 La Laguna, Tenerife, Spain 
\and
   Universe Cluster, Technische Universit\"{a}t M\"{u}nchen, Boltzmannstra\ss e 2, D-85748, Garching, Germany 
\and
   ARIES, Manora Peak, Nainital, Uttaranchal, India, 263129 
\and
   INAF/IASF Bologna, Via Gobetti 101, I--40129 Bologna, Italy 
\and
   University of Rochester, Department of Physics and Astronomy, Rochester, NY 14627-0171, USA 
\and
   Department of Physics and Astronomy, Clemson University, Clemson, SC 29634, USA 
\and
   Institute of Astronomy, University of Cambridge, Madingley Road, CB3 0HA, Cambridge, UK 
\and
   School of Physics, University College Dublin, Dublin 4, Republic of Ireland 
\and
   University of Alabama in Huntsville, NSSTC, 320 Sparkman Drive, Huntsville, AL 35805, USA 
\and
   Physics Department, University of California at Santa Barbara, 2233B Broida Hall, Santa Barbara, CA 93106, USA 
\and
   Institute of Physics, E\"otv\"os University, P\'azm\'any P. s. 1/A, 1117 Budapest, Hungary 
\and 
   CRESST and the Observational Cosmology Laboratory, NASA/GSFC, Greenbelt, MD 20771, USA 
\and
   Department of Astronomy, University of Maryland, College Park, MD 20742, USA 
}

\date{Received: 1 July 2010; accepted 8 February 2011}
 
\authorrunning{Rossi et al.}
\titlerunning{GRB 080928}

\abstract{We present the results of a comprehensive study  of the gamma-ray
  burst 080928 and of its afterglow. \object{GRB 080928} was a long burst detected by
  \swift/BAT and \fermi/GBM. It is one of the exceptional cases where optical
  emission had already been detected when the GRB itself was still radiating in the
  gamma-ray band. For nearly 100 seconds simultaneous  optical, X-ray and
  gamma-ray data provide a coverage of the spectral energy distribution of the
  transient source from about 1 eV to 150 keV.  In particular, we show that
  the SED during the main prompt emission phase agrees with
  synchrotron radiation. We constructed the optical/near-infrared light curve
  and the spectral energy distribution based on \swift/UVOT, ROTSE-IIIa
  (Australia), and GROND  (La Silla) data and compared it to the X-ray light
  curve retrieved from the \swift/XRT repository. We show that its bumpy shape
  can be modeled by multiple energy-injections into the forward shock.
 Furthermore, we investigate whether the temporal and spectral evolution of
  the tail emission of the first strong flare seen in the early X-ray 
  light curve can be explained by large-angle emission (LAE). We find 
  that a nonstandard LAE model is required to explain the observations.
 Finally, we report on the results of our search for
  the GRB host galaxy, for which only a deep upper limit can be provided.}


\maketitle


\section{Introduction \label{Intro}}

Currently there is a golden age in gamma-ray burst (GRB) research. The
dedicated \swift\ gamma-ray satellite was successfully launched in November
2004 (\citealt{Gehrels2004}), and has been in continuous operation for more than five
years now. Its sophisticated Burst Alert Telescope (BAT;
\citealt{Barthelmy2005a}), covering 15 to 150 keV, detects about 100 GRBs per
year with 3 arcmin localization accuracy (see J. Greiner's Internet page at
\texttt{http://www.mpe.mpg.de/$^\sim$jcg/grbgen.html}). In addition, about
once a month the European INTEGRAL gamma-ray satellite
(\citealt{Winkler2003}), usually pointing towards pre-planned
targets for days or weeks, localizes a GRB with similar position
accuracy (see \citealt{Vianello2009}). Also the Italian AGILE high-energy
satellite (\citealt{Tavani2009}) contributes about a handful of burst
detections and localizations per year 
(e.g., \citealt{Giuliani2008,Rossi2008}). Thanks to
\swift's rapid and autonomous slewing capabilities, in combination with its
highly sensitive X-ray telescope (XRT; \citealt{Burrows2005a}) as well as its
optical/UV telescope (UVOT; \citealt{Roming2005}), about 50 to 70
GRB optical afterglows can be localized annually, with 30 to 40 having redshifts
determined.

Roughly four years after \swift's launch  the \fermi\ Gamma-Ray Space
Telescope was launched into orbit (June 2008). Its Gamma-Ray Burst Monitor
(GBM; \citealt{Meegan2009}) and Large Area Telescope (LAT;
\citealt{Atwood2009}) cover an unprecedentedly wide energy range from 8 keV to
300~GeV. Up to the end of November 2010, LAT had localized 17 GRBs to positions of
less than a degree in error, of these, eight have optical afterglows and
redshifts\footnote{\texttt{http://fermi.gsfc.nasa.gov/ssc/observations/types/ grbs/grb\_table/}}.
Furthermore, a larger number of \swift\ GRBs
have also been detected  by \fermi/GBM, allowing a more thorough investigation
of the prompt emission above 150~keV.

Here we report on the analysis of the prompt gamma-ray emission and the
afterglow of \object{GRB 080928}, as well as on the search for its  host galaxy. This
burst was detected by \swift/BAT and \fermi/GBM but not seen by \fermi/LAT.
Its afterglow was rapidly found, and \cite{Vreeswijk2008a} report a redshift
of $z=1.692$. The burst is of particular interest since both optical and
X-ray emission was detected by \swift/UVOT and \swift/XRT, respectively, when
the GRB was still radiating in the gamma-ray band. This makes it one of a rare
number of cases (e.g., GRBs 041219A, 050820A, 051111, 061121;
\citealt{Shen2009}), where a broad-band spectral energy distribution (SED)
from about 1~eV to 150~keV can be constructed for the prompt emission phase.

Throughout this paper we adopt a world model with $H_0=71$ km s$^{-1}$
Mpc$^{-1}, \Omega_{\rm M}=0.27, \Omega_\Lambda=0.73$ 
\citep{Spergel2003}. For the flux density of the afterglow we use 
the usual convention $F_\nu(t) \propto t^{-\alpha} \nu^{-\beta}$.

\begin{figure*}[th!]
\begin{center}
\includegraphics[width=18.4cm,angle=0]{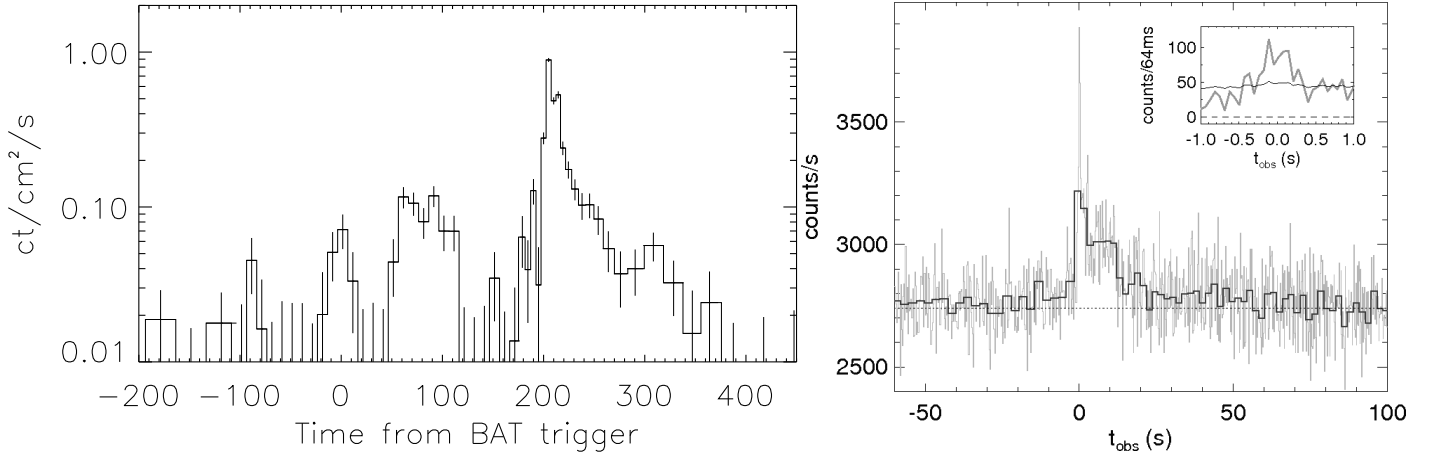}
\caption{{\it Left}: The light curve of \object{GRB 080928} as seen by \swift/BAT. 
\swift\ triggered at the gamma-ray peak at $t_0=0$, which was followed
by at least two more peaks with the maximum at $t_0+204$~s. 
There may be a faint precursor of the main burst at $t_0-90$~s. 
{\it Right}: \fermi/GBM light curve  of
the NaI detectors \#0, \#3, \#4, and \#7 combined with $2$~s resolution (black
line) and $0.256$~s resolution (gray line). A zoom into the $64$~ms-binned,
background-subtracted light curve around the peak is shown in the
inset. Variability on time scales of $\sim128$~ms is detected at $3\sigma$
(solid gray line) above the background plus shot noise fluctuations. 
In this figure the time zero-point is the \fermi/GBM trigger time 
$t_{0, \rm GBM}$ (Eq.~\ref{t0}).}
\label{fig:gbm_lc}
\end{center}
\end{figure*}

\section{Data and analysis}

\subsection{Swift/BAT and Fermi/GBM data}
\label{History}

The long-burst \object{GRB 080928} triggered the Burst Alert  Telescope of \swift\ at
$t_0$ = 15:01:32.86  UT \citep{Sakamoto08GCN8292} on the 28 of September
2008. This was an image trigger lasting 112 seconds. The prompt emission
detected in the BAT began with a faint precursor at $t_0-90$~s, then weak
emission starting at $t_0-20~s$ and lasting for $40$~s, followed by a second,
slightly brighter peak starting at $50$~s and ending at $120$~s after the
trigger (Fig.~\ref{fig:gbm_lc}).  The main emission of the GRB started at
$t_0+170$ seconds, with two peaks at $204$ and $215$ seconds\footnote{If not
  stated otherwise, for the rest of the  paper all times refer to the
  zero-point $t_0$.}. Another less significant peak is detected around 310~s
before fading out to at least $400$ seconds when \swift\ had to stop observing
due to its entry into the South Atlantic Anomaly (SAA) and the noise level
became too strong for any late emission to be detected in the BAT
\citep{Cummings2008a,Fenimore2008GCN8297, Sakamoto08GCN8292}. 

The main burst emission also triggered the Gamma-Ray Burst Monitor onboard
\fermi\ \citep{Paciesas2008a}, while the INTEGRAL satellite was passing
through the SAA during the time of \object{GRB 080928} and thus could not observe the
burst with the anti-coincidence shield of the spectrometer SPI (SPI-ACS,
\citealt{Rau2005}). GBM consists of 12 sodium iodide (NaI)
detectors that cover the energy band between 8 keV and 1 MeV and two bismuth
germanate (BGO) scintillators that are sensitive at energies between 150\,keV
and 40\,MeV. Emission from the burst was predominately seen in the NaI
detectors. The GBM light curve (Fig.~\ref{fig:gbm_lc}) shows a single pulse
corresponding to the emission maximum observed by \swift \ at
\begin{equation}
t_{0, \rm GBM}=t_0+204\ \mbox{s}\,. 
\label{t0}
\end{equation}

We analyzed data collected by BAT between $t_0-239$~s and $t_0+494$~s
in event mode with 100~$\mu$s time resolution and about 6 keV  energy
resolution. The data were processed using standard BAT  analysis tools,
and a background-subtracted light curve was produced  using the tool
\texttt{batmaskwtevt} with the best source position.  For spectral analysis,
the data were binned so that the signal-to-noise ratio was at  least $3.0$.
During the main peak, the bin edges were chosen to match the \swift/XRT spectral
bins.  The spectra were fit using \texttt{Xspec v12.5.0}.

The spectral analysis of the \emph{Fermi} data  was performed with the
software package  \texttt{RMFIT v3.2rc1} using Castor statistics. Here, we
analyzed the GBM spectra of the brightest four NaI detectors (\#0, \#3, \#4 \&
\#7) for two different integration windows, one covering the broad emission
maximum from $t_{0, \rm GBM}-5.248$~s to $t_{0, \rm GBM}+24.448$~s, while the
second was constrained to $\approx4$~s around the peak ($t_{0, \rm
 GBM}-1.152$~s to $t_{0, \rm GBM}+2.944$~s). The variable GBM background was
subtracted for all detectors individually by fitting an energy-dependent,
third-order polynomial to the background data.  The background interval used for
the analysis was  from $t_{0, \rm GBM}-100$~s to $t_{0, \rm GBM}-50$~s and
from  $t_{0, \rm GBM}+100$~s to $t_{0, \rm GBM}+350$~s. We used the standard
128 energy bins of the CSPEC data-type, using the channels above 8 keV of the
NaIs and ignoring the so-called overflow channels.

\subsection{Swift/XRT data \label{XAG} }

\swift/XRT started to observe the BAT GRB error circle 170 seconds
after the trigger and found an unknown X-ray source at coordinates R.A.
(J2000)= $6^{\rm h}20^{\rm m}16\fs 87$, Dec. = $-55^\circ11'58\farcs5$, with
a final uncertainty of 1.4 arcsec \citep{Osborne2008a,
Sakamoto08GCN8292}. Observations continued until 2.7 days after the GRB,
when the source became too faint to be detected.

We obtained the X-ray data  from the \swift \ data archive and the light curve
from the \swift \ light curve repository \citep{Evans2007a, Evans2009a}. To
reduce the data, the software package \texttt{HeaSoft} 6.6.1 was used\footnote{
\texttt{http://heasarc.gsfc.nasa.gov/docs/software/lheasoft}} with
the calibration file version
\texttt{v011}\footnote{\texttt{heasarc.gsfc.nasa.gov/docs/heasarc/caldb/swift}}.
Data analysis was performed following the procedures described in Nousek et
al. (2006). We found that  the X-ray emission was only bright enough to
perform a spectral analysis in the first two observing blocks
(\texttt{000}--\texttt{001}). However, the early windowed timing
(\texttt{wt}) mode and photon counting (\texttt{pc}) mode data were highly
affected by pile-up. To account for this effect, we applied the methods
presented in \citet{Romano2006a} and \citet{Vaughan2006a}.

Owing to the brightness of the source in \texttt{wt} mode, a time filter was
defined to have at least 500 counts (background-subtracted) for every
spectrum. In \texttt{pc} mode the average number of counts per spectrum is 300
due to pile-up. On these spectra $\chi^2$-statistics were applied. 
Observing block \texttt{001} has only 102 counts 
(background-subtracted), so we could only apply \emph{Cash}
statistics \citep{Cash1979a, Evans2009a}.
In total, from both observing blocks
we extracted the SED for 27 epochs, covering 1.4 days.

Following \citet{Butler2007a}, 
we initially fitted the \texttt{pc}-mode spectra with an absorbed power-law 
to obtain $N_{\rm H}^{\rm host}$ using \texttt{Xspec v12.5.0}. This model
consists of two absorption  components, one in the host frame 
and another one in the Galaxy. For both
absorbers we used the T\"ubingen abundance template by
\citet{Wilms2000a}, with the Galactic absorption fixed to $N^{\rm Gal} _{\rm
H} = 0.56\times 10^{21}\,\rm{cm}^{-2}$ \citep{Kalberla2005a}.
The spectra were then fitted in two steps. First, all \texttt{pc}-mode spectra
of the XRT observing block \texttt{000} were stacked using the FTOOL
\texttt{mathpha}
\citep{Blackburn1995}\footnote{\texttt{http://heasarc.gsfc.nasa.gov/ftools/}}. 
This spectrum contained about 1000 counts. The fitted
absorbed power law is characterized by a spectral  slope 
of $\beta_{\rm X} = 1.09^{+0.07}
_{-0.10}$ and an effective hydrogen column density of $N_{\rm H}^{\rm host} =
3.6 ^{+1.8} _{-2.2} \times 10^{21}\,\rm{cm}^{-2}$.  The spectral slope agrees
with the observed mean value of $\beta_{\rm X} \sim1$ found by, e.g., 
\citet{Racusin2009a} and \citet{Evans2009a}. Having derived $N_{\rm H}^{\rm host}$ in this way, the early spectra
(\texttt{wt}-data) were fitted with an absorbed power law in which
$N_{\rm H}^{\rm host}$  was fixed to the previously derived value.

\subsection{Optical/NIR data \label{sec:optnir}}

\swift/UVOT started observing about 3
min after the trigger, still before the onset of the main emission of the
GRB, and immediately found an optical afterglow candidate 
\citep{Kuin2008_GCN8298,Sakamoto08GCN8292}.  The redshift reported by
\citet{Vreeswijk2008a} was later refined to $z=1.6919$ by
\cite{Fynbo2009a}.\footnote{For 
this redshift the distance modulus is $m-M=45.54$ mag, the
luminosity distance $3.95\,\times\,10^{28}$ cm, the look-back
time 9.76 Gyr (3.91 Gyr after the Big Bang), and 1 arcsec on the sky
corresponds to a projected distance of 8.56 kpc.}

\swift/UVOT data were analyzed using the standard analysis software
distributed within FTOOLS, version 6.5.1. For all the detections, the source
count rates were extracted within a 3\arcsec\ aperture. An aperture correction
was estimated from selected nearby point sources in each exposure and applied
to obtain the standard UVOT photometry calibrated for a 5\arcsec\ aperture.

Ground-based follow-up observations were performed by our group using the
ROTSE-IIIa 0.45m telescope in Australia \citep{Rykoff2008_GCN8293} and the MPG/ESO
2.2m telescope on La Silla, Chile, equipped with the multichannel imager
GROND \citep{Greiner2007a, Greiner2008a}. This data set
(Tables~\ref{tab:magsROTSE}, \ref{tab:magsUVOT},
\ref{tab:mags}; Fig.~\ref{fig:lc_multicolorAlex})
was supplemented by data
published from the VLT \citep{Vreeswijk2008a, Fynbo2009a}, and the $16''$ 
Watcher telescope in South Africa \citep{Ferrero08_GCN8303}.

\begin{figure}[htb]
\includegraphics[width=8.8cm]{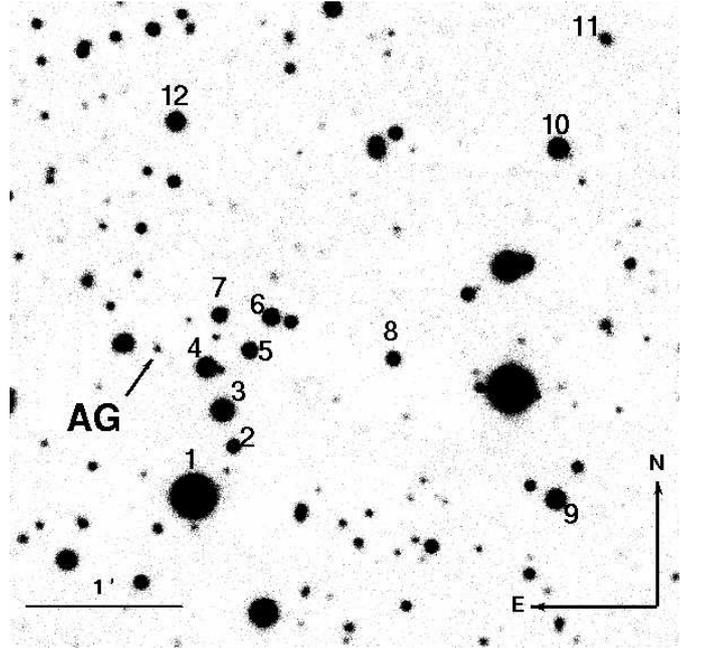}
\caption{Finding chart of the afterglow of \object{GRB 080928} (GROND $i^\prime$ band, 
at $0.603$ days after the burst). The afterglow (AG) 
and the secondary photometric
standards used (Table~\ref{tab:std}) are indicated.}
\label{fig:finding}
\end{figure}

ROTSE-IIIa data were analyzed with a PSF photometry package based on
DAOPHOT  following the procedure described in \citet{Quimby2006a}.
GROND optical/NIR data were analyzed through standard PSF photometry using
DAOPHOT tasks under IRAF \citep{Tody1993} similar to the procedure
described in \citet{Kruhler2008a}. Aperture photometry was applied
when analyzing the field galaxies, using the DAOPHOT package
\citep{Warmels1992}.  Afterglow coordinates were derived from the GROND 3rd
epoch  $g^\prime r^\prime i^\prime z^\prime $-band data. The stacked image has
an astrometric precision of about 0.3 arcsec, corresponding to the RMS accuracy
of the USNO-B1 catalogue (Monet et al. 2003). The coordinates of the optical
afterglow (Fig.~\ref{fig:finding}) are R.A. (J2000)= $06^{\rm h}20^{\rm
  m}16\fs 83$, Dec. =$-55^\circ11'58\farcs9$ (Galactic coordinates $l$, $b$ =
$263\fdg 82\,,-26\fdg 31$).  Magnitudes were corrected for Galactic
extinction using the interstellar extinction curve derived by
\citet{Cardelli1989} and by assuming $E(B-V)=0.07$ mag \citep{Schlegel1998}
and a ratio of total-to-selective extinction of $R_V=3.1$.

\begin{table*}[t]
\begin{minipage}{\textwidth}
\renewcommand{\footnoterule}{}
\centering
\begin{footnotesize}
\caption{
Spectral fit results for \swift/BAT and the \fermi/GBM NaI detectors
\#0,3,4,7.}
\label{tab:gbm_spec_one}
\begin{tabular}{l|l|cccccc|cc}
\toprule
instrument  & model  & $\tilde{\beta_1}$                  & $E_1$                & $\tilde{\beta_2}$                        & $E_2$                        & $\tilde{\beta_3}$      & $\chi^2/$d.o.f.  & $F_{\rm ph}$ (0.3-1) & $F_{\rm ph}$ (0.3-10) \\  
\midrule
\multicolumn{8}{c}{ \bf{ 46.5~s $< t_0 < $ 121.0~s }}\\[1mm]
BAT-GBM     & db-pl    & --                     & $ 12.37_{-12.37}^{+1.52}$ & $ 1.92_{-0.18}^{+0.13}$      & $ 143_{-64}^{+37}$ & --                     & 582 / 560       & --                            & --                            \\[3mm]
\multicolumn{8}{c}{ \bf{ 202.848~s $< t_0 <$ 206.944~s   = --1.152\,s$<t_{0, \rm GBM}<$2.944\,s}}\\[1mm]
GBM         & s-pl    & --                     & --                       & $1.75\pm0.04$             & --               & --                     & 422 / 438        & $0.131\pm0.004$               & $0.21\pm0.06$                 \\
GBM         & Band   & --                     & --                       & $1.24\pm0.16$             & $108\pm24$       & $3.3\pm4.6$            & 411 / 436        & $0.07\pm0.02$                 & $0.09\pm0.02$                 \\
XRT-BAT-GBM & db-pl    & $0.62_{-0.18}^{+0.10}$   & $3.94_{-0.62}^{+0.56}$     & $1.74_{-0.08}^{+0.05}$       & $131_{-16}^{+6}$   & --                   &  639 / 581       & --                            & --                            \\[3mm]
\multicolumn{8}{c}{ \bf{ 198.752~s $< t_0 <$ 228.448~s = --5.248\,s$<t_{0, \rm GBM}<$24.448\,s}}\\[1mm]
GBM         & s-pl    & --                     & --                       & $1.90\pm0.04$             & --              & --                     & 571 / 438        & $0.035\pm0.002$               & $0.051\pm0.002$               \\
GBM         & Band   & --                     & --                       & $1.51\pm0.16$             & $70\pm17$        & $2.5\pm0.7$            & 564 / 436        & $0.023\pm0.036$               & $0.032\pm0.049$               \\
XRT-BAT-GBM & db-pl    & $1.14\pm0.03$   &--                        & $ 1.81\pm0.05$      & $ 132_{-16}^{+49}$ & -- & 643 / 674        & --                            & --                            \\
\bottomrule
\end{tabular}
\tablefoot{ Column 2: s-pl stands for single power law SED, 
db-pl for a double broken power law, and Band for a Band function.
Columns 3 to 8: Results of the fit.
Columns 9 and 10: The 
photon flux $F_{\rm ph}$ [ph/cm$^2$/s] in the high-energy domain from 0.3 to 1
MeV and 0.3 to 10 MeV, respectively, extrapolated from the GBM data.
All other energies are given in units of keV.}
\end{footnotesize}
\end{minipage}
\end{table*}

During our first two epochs of GROND observations \citep{Rossi08_GCN8296} the
weather conditions were not good, with the seeing always higher than 2.5 arcsec
and strong winds ($>10$ m/s).  Therefore, it was not possible to separate the
afterglow from a nearby galaxy that first became separately visible on the 
third-epoch images (seeing 1.5 arcsec; see Sect. \ref{hosts}).  To correct
for the contribution of this galaxy, we performed image subtraction using the
\texttt{HOTPANTS}
package\footnote{http://www.astro.washington.edu/users/becker/hotpants.html}. 
We applied image subtraction on the first, second, and third epoch GROND
images, using the fifth GROND epoch images as a template. This gave good
results for all bands except $g^\prime$,  which is affected by a low-quality
point spread function. Therefore, for this band  we performed a simple
subtraction of the flux of the galaxy component, with the flux derived from
the fifth-epoch images. 
Calibration of the field in $JHK_S$ was performed
using 2MASS stars (Table~\ref{tab:std}). 
The magnitudes of the selected stars were 
transformed into the GROND filter system and finally into AB magnitudes
using $J(AB) = J(Vega) + 0.91$, $H(AB) = H(Vega) + 1.38$, $K_s(AB) =
K_s(Vega) + 1.79 $ \citep{Greiner2008a}.

Watcher data \citep{Ferrero08_GCN8303}, VLT data \citep{Vreeswijk2008a,
Fynbo2009a}, and ROTSE-IIIa data were calibrated using USNO-B1 field
stars. In order to take these different calibrations into account, we compared
the $r^\prime$-band photometry of the GROND  secondary standard stars with the
corresponding $R$-band magnitudes from USNO-B1. In doing so, we obtained a
correction of 0.40$\pm$0.15 mag for USNO-B1. After shifting these afterglow
data to the GROND $r^\prime$ band, we finally subtracted the GROND fifth-epoch
flux of the galaxy  closest to the afterglow (see Sect.~\ref{hosts}) from the
Watcher and VLT observed magnitudes, which shifted the afterglow magnitude by
+0.05 mag and +0.11 mag, respectively. The correction for the ROTSE-IIIa data was
even smaller and, therefore, set to zero. The complete data set is shown in
Fig.  \ref{fig:lc_multicolorAlex}.

\section{Results and discussion}

\subsection{The prompt emission phase}

\begin{figure}[htb]
\centering
\includegraphics[width=8.9cm]{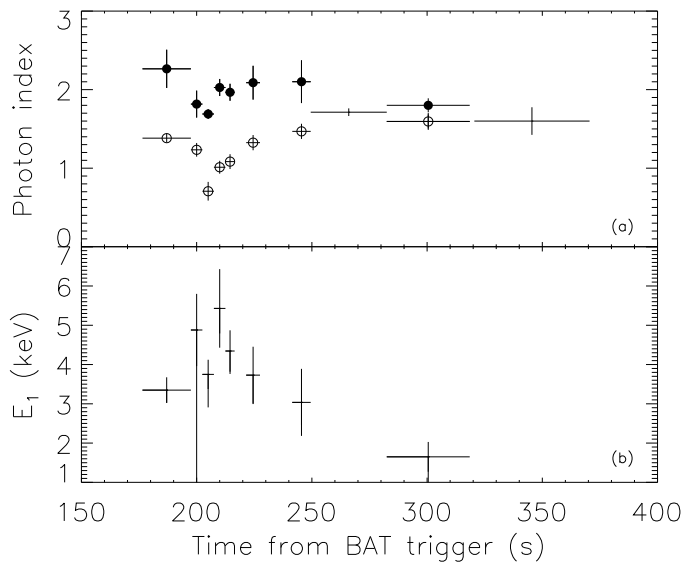}
\caption{
Spectral parameters of the prompt emission using the time-resolved XRT-BAT-GBM
data. {\it (a):} The evolution of the photon index from fits to BAT-GBM and XRT
data.  Open circles show the low-energy index $\tilde{\beta_1}$ below the
break energy $E_1$  of a single broken power law and the filled circles
represent the  high-energy index $\tilde{\beta_2}$ above $E_1$.   Points with
no plot symbols (error bars only) are the best-fit results using only a simple power law. 
{\it (b):} The low-energy break energy,
$E_1$, from fits to the BAT and XRT data. During the flare at $208$~s 
spectral evolution is seen, similar to what was also
detected in other afterglows (e.g., \citealt{Falcone2007a}).}
\label{fig:ebreak}
\end{figure}


The prompt gamma-ray emission is dominated by a strong peak starting at
$170$~s, which reached its maximum at $204$~s and was detected by GBM, BAT,
and XRT. In addition, XRT also detected a  second weaker peak
at $357$~s.  The first peak and the main peak were also detected by UVOT in the 
$white$ and $v$ bands. 

\subsubsection{From gamma-rays to X-rays \label{sec:prompt}}

During the first peak of the prompt emission (in the interval $t_0-23.5$~s$ <
t < t_0+16.5$\ s) we could fit only a simple power law  to the BAT data with a
photon  index $1.67\pm0.34$.  We also fitted the BAT-GBM data during the
second peak ($t_0+46.5$~s$ < t < t_0+121$\ s) and the XRT-BAT-GBM data
during the main peak ($t_0+198.75$~s$ < t < t_0+228.4$\ s). For both peaks we
found a peak energy of $\approx 130$\ keV, though we could not constrain the
index above the peak (Table~\ref{tab:gbm_spec_one}). No spectral analysis was
possible for the precursor. 

For the GBM-only data, two different empirical models were applied to fit the
spectra: a simple power law and a Band function \citep{Band1993}, which
smoothly connects two power laws. The burst
was faint for the GBM, especially at energies above 150\,keV. Thus, the more
complex model of a Band function could not be constrained sufficiently and the
simple power law is preferred for both time intervals.  

Table~\ref{tab:gbm_spec_one} summarizes the fits of the SED for  the
XRT-BAT-GBM data for two time intervals around the main peak in the gamma-ray
light curve.  In particular, we performed a spectral fit for the peak centered
around 204~s.  For joint fits with BAT and XRT, we used an absorbed power law
with the Galactic and the GRB host column densities fixed to the values found
in Sect~\ref{XAG}. 

Figure~\ref{fig:ebreak} shows the time evolution of the SED in the BAT band and
the joint BAT-XRT band during the first $400$~s after the BAT trigger. 
For the
three early peaks in the BAT light curve (Fig.~\ref{fig:gbm_lc}) the error
bars are too large to indicate any spectral evolution. During the main
gamma-ray peak at $204$~s, however, there is evidence of a spectral softening
when the peak is developing and a spectral hardening after the peak. 
After the light curve peak, the situation is reversed. 
This behavior is similar to what has been  found for
\object{GRB 060714} \citep{Krimm2007a}. Also, the power law indices, as well as  the
break energy, are consistent with the corresponding values found in gamma-ray
flares \citep{Krimm2007a}.

In the cases
where a broken power law model is the best fit ($\Delta\chi^2 > 4$), the break
energy, as well as the high-energy index and the low-energy index, is well
constrained, so essentially BAT is fitting the high-energy index, XRT is
fitting the low-energy index, and the joint fit fits an  average index,
becoming dominated by the low-energy emission where the  BAT statistics are
poor. Remarkably, even though the break energy is  always between 1 and 5 keV,
i.e. well below the BAT and GBM window, the prompt emission
flare is still very bright in BAT and GBM. Moreover, it is ten times brighter
than the peak on which BAT triggered.

\subsubsection{From gamma-rays to the optical \label{sec:joint}}

\begin{figure*}[t]
\centering
\includegraphics[width=18.9cm,bb=40 20 650 610, angle=0,clip]{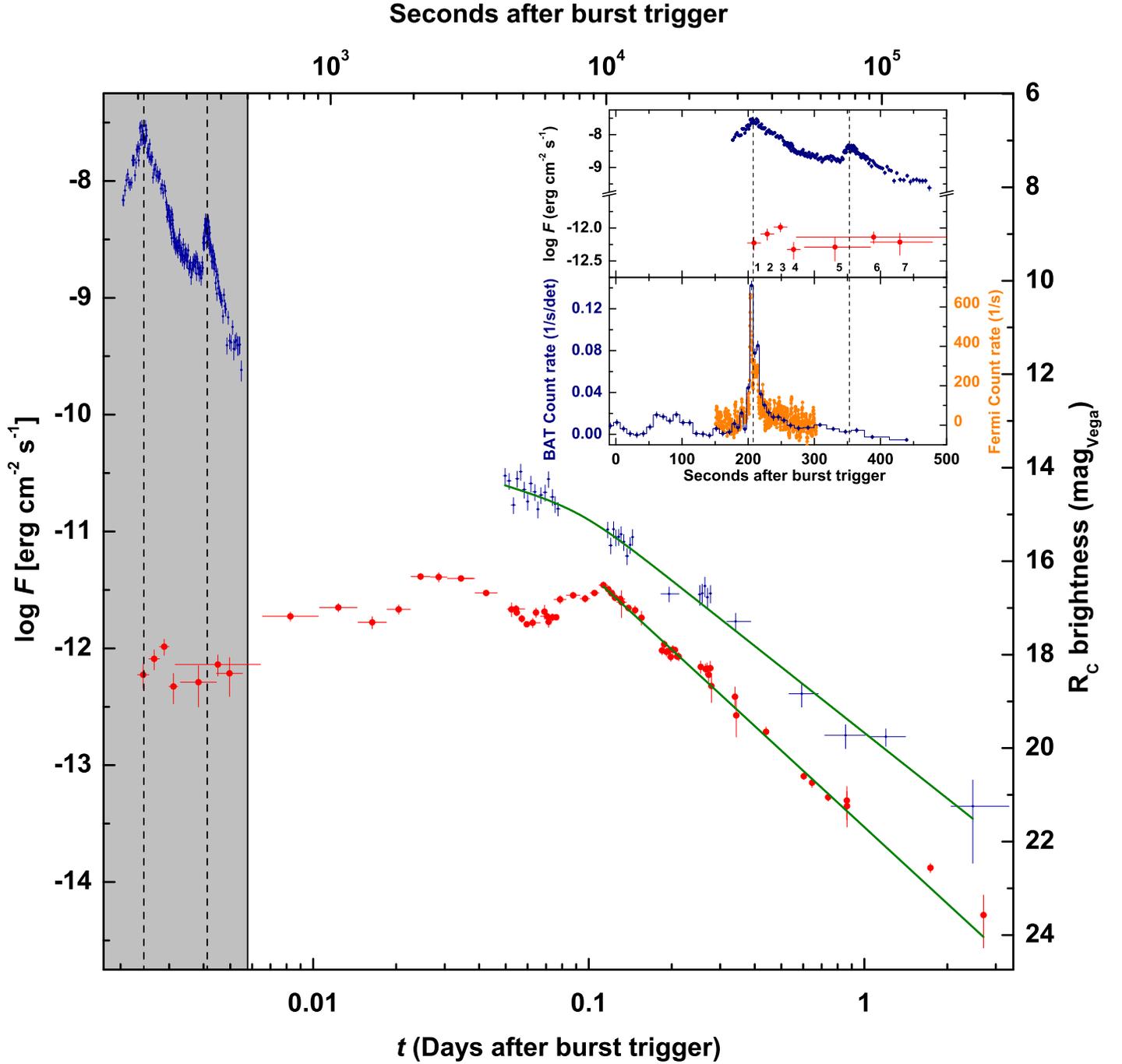}
\caption{
Temporal evolution of the optical (composite light curve with all data 
shifted to the $R_c$ band) and X-ray afterglow (0.3 to 10 keV) of GRB
080928 (optical: red circles, X-ray: blue
error bars).  The upper limits are not shown here to avoid confusion.
 The zoom-in shows the early phase (also highlighted in gray 
in the big figure) where it is compared with the BAT-GBM prompt emission. 
The dashed vertical lines indicate the peak times of the two X-ray
flares. The curve represents the best fit of the late-time data.}
\label{fig:lcspec}
\end{figure*}

\object{GRB 080928} is one of those exceptional cases where optical and X-ray data
could be obtained while the source was still being detected in the gamma-ray
band (Fig.~\ref{fig:lcspec}). The analysis of the joint UVOT-XRT-BAT-GBM SED
allows us to follow the evolution of the prompt emission during all the main
flaring activity observed between $199$ and $557$ seconds after the trigger
from 1 eV to 150 keV.

The prompt gamma-ray emission detected by BAT and GBM is
dominated by the strong peak at $204$~s.  Possibly  physically related to that
is a strong peak in the X-ray emission seen by XRT about four seconds
later at $208$~s, which was followed by a less intense X-ray peak at
$357$~s. The latter has no obvious counterpart in the gamma-ray emission. 
The optical light curve monitored by UVOT shows a first peak at $249\pm10$~s,
i.e. 45 seconds after the main peak of the prompt emission and $41$~s after
the main peak in the X-ray flux.  

To gain deeper insight into the early emission properties and on
their time evolution, we then included the optical data and constructed the SED
from the optical to the gamma-ray band for six time intervals defined by the
first six optical detections by UVOT,
starting at $199$~s and finishing $479$~s after the trigger (Table~\ref{tab:joint},
Fig.~\ref{fig:lcspec}). In doing so, we exclude the sixth optical measurement (ROTSE-IIIa)
because it covers a rather big time interval.

\begin{figure}[htb]
\includegraphics[width=8.8cm]{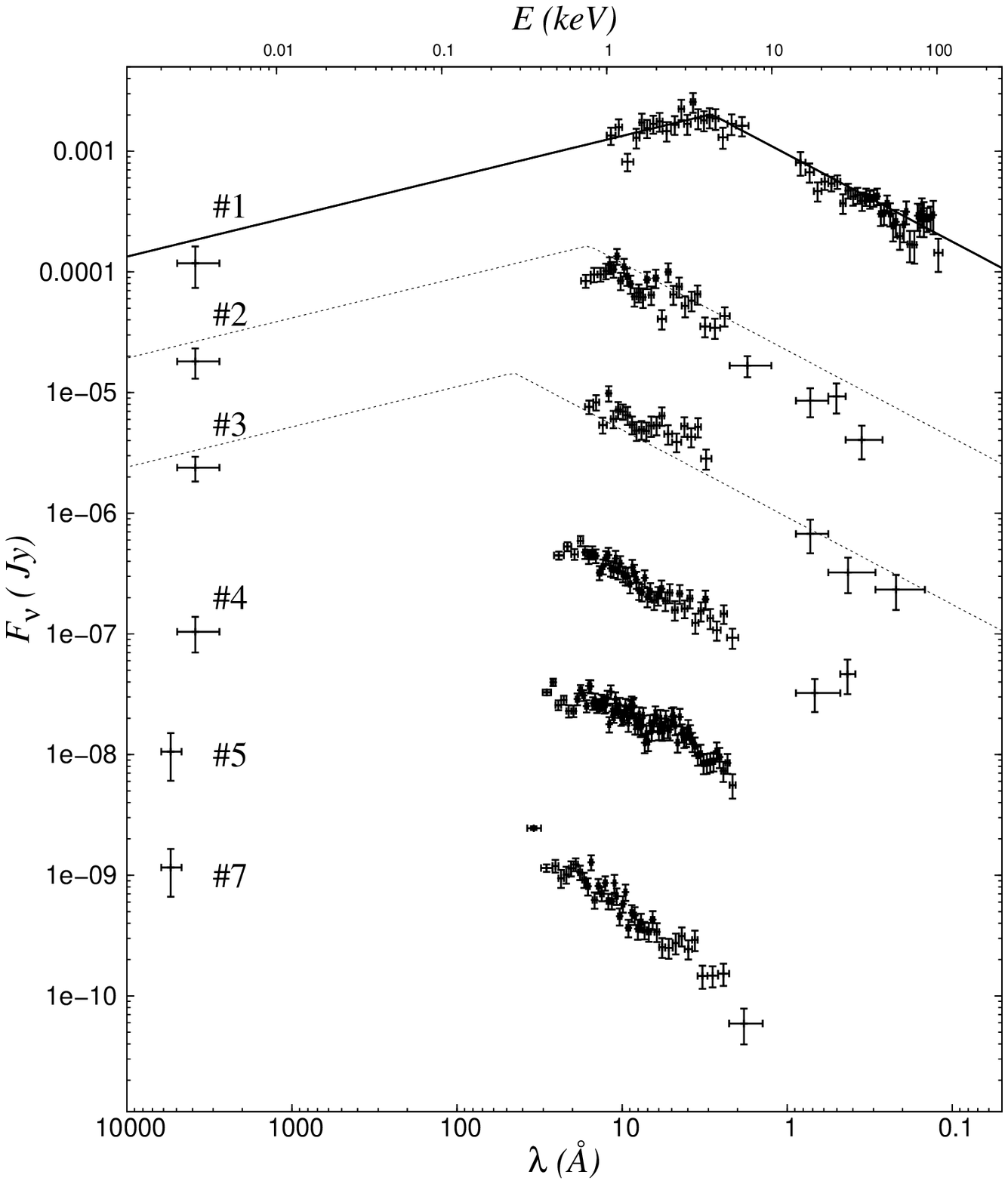}
\caption{The spectral energy distribution of the combined early emission
during the time when the first six optical data points were obtained by
\swift/UVOT \emph{white} and \emph{v} filters. 
The corresponding time intervals are listed in
Table~\ref{tab:joint}. The fluxes of the curves \#2, 3, 4, 5, and 7 have
been multiplied for clarity by $10^{-1}$, $10^{-2}$, $10^{-3}$, $10^{-4}$,
and $10^{-5}$, respectively. The fits for \#2, 3 were 
obtained by fixing the high-energy slope to the corresponding slope
obtained for SED \#1, the low-energy slope to 1/3, and by matching the 
expected break energy following the nonstandard LAE model (Sect.~\ref{sec:largeangle}).}
\label{fig:peaksed2}
\end{figure}

During the first five time intervals, BAT and GBM were still detecting
gamma-ray emission (the main gamma-ray peak occurred when UVOT was already
observing), while during the last two time intervals the fluence in the
gamma-ray band was too low  to constrain the spectral properties.
Figure~\ref{fig:peaksed2} shows the fit to the data  from about 1 eV to up to
150 keV. In the following, we first focus on SED \#1.  Here, we fit the data
with a broken power law with the X-ray data corrected for Galactic and
GRB host absorption (see Sect.~\ref{betaXX}) and the optical data corrected
for the Galactic and GRB host extinction. 

For the time interval \#1 (Table~\ref{tab:joint}), we combined the first
optical UVOT detection (Table~\ref{tab:magsUVOT}) with the XRT and the BAT-GBM
detection from $202.8$~s to $206.9$~s. A sharp break is clearly visible at an
energy around 5 keV. For SED \#1 the soft X-ray data, $E<1$ keV,
shows too much scatter and therefore could not be used for the analysis.
Assuming that SED \#1
represents the spectral energy distribution of the synchrotron light of a single
radiating component from about 1~eV to 150~keV (see also \citealt{Shen2009}),
we fitted the data with a broken power law while fixing the 
low-energy index to its theoretically expected 
value $\beta=-1/3$ (i.e., rising with energy). The slope of the high-energy index 
is then found to be $\beta=0.72\pm0.06$ ($\chi^2$/d.o.f.=$66.8$/75) 
with a spectral break at an energy of $4.30\pm 0.45$ keV.
The corresponding UVOT data point lies $1\sigma$ below the best fit 
(Fig.~\ref{fig:peaksed2}). 

If we identify the break in the SED as the position of the minimum injection
frequency $\nu_m$ of an ensemble of relativistic electrons in the slow cooling
regime  $(\nu_m < \nu_c$, with $\nu_c$ being the cooling frequency), then we
expect a low-energy spectral index of $-1/3$ and a high-energy spectral index
of $(p-1)/2$, where $p$ is the power law index of the electron distribution
function  ($N(\gamma) d\gamma \propto \gamma^{-p} d\gamma$). The measured
low-energy spectral index ($-0.39\pm0.06$) basically agrees with the
theoretically expected value. The measured high-energy spectral index is
$0.72\pm0.06$, leading to $p=2.44\pm0.12$, which is a reasonable value for
relativistic shocks, both theoretically (\citealt{Achterberg2001, Kirk2000})
and observationally (e.g., \citealt{Kann2006a, Starling2008,
Curran2010}). 

On the other hand, if the break is the cooling frequency in
the fast-cooling regime, then  we expect a low-energy spectral index of $-1/3$
and a high-energy spectral index of 0.5. Within errors, the latter disagrees
 with the observations, the spectral slope is $0.72\pm0.06$, and the
discrepancy is $3.7\sigma$. However, it is quite possible that the snapshot of
the high-energy part of the SED we observe in our time window is the average
of a rapidly evolving SED that accompanied the rapidly evolving light curve. 

Making the step to the SEDs \#2 to \#4, we are faced with the problem that the
break seen in SED \#1 is not detectable anymore, most likely because 
the peak energy $E_p$ has moved to lower energies. However, given that we see
a large flare in the X-ray light curve, part of the data
allow us to investigate if the evolution of the SED
is compatible with large-angle emission.

\begin{table}[htb]
\begin{minipage}{\columnwidth}
\renewcommand{\footnoterule}{}
\centering
\caption{Results of the joint optical to gamma-ray
spectral fit ($\sim$1 eV to $\sim$150 keV).}
\begin{tabular}{lccccc}
\toprule
\# & \multicolumn{2}{c}{Optical\hspace*{5mm}} &\multicolumn{2}{c}{XRT-BAT-GBM} &
$E_{\rm break}$ \\
&interval &time& interval &time& (keV) \\
\midrule
1 & 199.0 - 219.0 &208.7 & 202.8 - 206.9 & 204.8 &  $4.30(45) $\\
2 & 219.0 - 238.0 &228.7 & 227.5 - 234.5 & 231.0 &  $0.78(39) $\\
3 & 239.0 - 258.0 &248.7 & 241.5 - 249.5 & 245.5 &  $0.28(18) $\\
4 & 259.0 - 278.7 &268.7 & 259.0 - 278.7 & 268.7 &  -- \\
5 & 285.0 - 385.0 &331.3 & 318.5 - 372.5 & 344.4 &  -- \\
6 & 272.7 - 556.7 &389.6 & 272.7 - 477.9 & 361.0 &  -- \\
7 & 385.0 - 478.7 &429.3 & 385.0 - 477.9 & 428.9 &  -- \\
\bottomrule
\end{tabular}
\tablefoot{Columns 2 and 3: Seven time intervals (in units of seconds)
defined by the first seven optical data points (``epochs'';
Fig.~\ref{fig:lcspec}) and their logarithmic mean.
Column 4,5: The corresponding time spans when high-energy photons
were collected and their logarithmic mean.
Column 6: The break energy, including its $1\sigma$ error.
For further details see Sect.~\ref{sec:joint}.}
\label{tab:joint}
\end{minipage}
\end{table}

\subsubsection{Large-angle emission \label{sec:largeangle}}

X-ray flares are commonly observed in GRB afterglows, with the most
prominent example beeing \object{GRB 050502B}
(e.g., \citealt{Chincarini2007,Chincarini2010,Burrows2005b}).  The early flares of GRB
080928 are among the strongest flares seen so far.  While much stronger flares
have been observed (GRBs 060124, \citealt{Romano2006a}; 061121, \citealt{Page2007a}), the first flare seen in the
afterglow of \object{GRB 080928} is even stronger in terms of peak count rate than  the
flare of \object{GRB 050502B}. In particular, it has good enough data to investigate whether
its radiation tail can be interpreted as large-angle emission
(LAE; \citealt{Fenimore1997,Kumar2000a}). 

Figure~\ref{fig:lcspec} shows that between epochs \#3 and \#4 the optical light
curve  is falling, while thereafter it remains constant within the errors.
The figure also shows that  after the fifth optical epoch the X-ray light curve
has a second flare.  
We wish to study only the interval when the light curve has a constant power law index,
and therefore we only include the first three data points in 
Table~\ref{tab:joint} in our analysis.  
In doing so, we fixed the value for the spectral slopes to the one
for SED \#1 ($\beta=-1/3$ for the low-energy part as given by synchrotron theory and
$0.72$ for the high-energy part as it follows from the fit).

Within the standard LAE model, there is a one-to-one correspondence between 
photon
arrival time $t$ and location of emitting fluid: $t=(1+z) \, r \theta^2/c$,
where $r$ is the source radius and $\theta$ the direction of fluid motion
relative to the line that connects the center of the explosion 
and the observer.  So, the observer receives emission from
fluid regions moving at progressively larger angles $\theta$. Thus, at
different times, the observer receives emission from different regions and
from different electrons. Thereby, the  following assumptions are made: (1)
the electron population is the same at all angles $\theta$ and (2) the surface
brightness of the emitting shell is uniform in angle.  From these assumptions,
it follows that the flux decreases as $t^{-(2+\beta)}$.  From
the first assumption, it follows that the peak energy should decrease as
$t^{-1}$. In the $\nu^{1/3}$ part of the spectrum, the optical LAE should
then decay as $t^{-5/3}$, however, our data show that the optical flux is
rising between epochs 1 and 3 (Fig.~\ref{fig:lcspec}).

If the entire emission between the first and the second X-ray flares  is of
LAE origin, then the fact that the optical flux increases at epochs 2 and 3
(instead of decreasing as $t^{-5/3}$), while the X-ray flux decreases, implies
that the aforementioned  assumption (1) of the LAE model is incorrect. In
particular, it implies that $E_p$ for the electrons at larger angles
(corresponding to epoch 3) is lower than at smaller angles (corresponding to
epoch 1), at the same lab-frame time. In other words, the rising optical flux is
compatible with the LAE interpretation only if $E_p$ decreases with observer
time faster than $t^{-1}$. 

Therefore, we applied a non standard LAE model. We assumed that the local
synchrotron peak flux $F_p$ and the peak energy $E_p$ depend on the
viewing angle $\theta$. In doing so, we make the ansatz that an observer
located at an angle $\theta$ relative to us would observe  a peak flux  and
peak energy  evolving as $F_p(\theta) \propto \theta^{-2a}$ and
$E_p \propto \theta^{-2b}$, respectively. The evolution of the measured peak
flux and  peak energy after  relativistic boosting is then $F_p \propto
(t-t_p)^{-2-a}$ and  $E_p \propto (t-t_p)^{-1-b}$, respectively, where $t_p$
is the unknown zero point.  The resulting LAE X-ray light curve  above the
peak energy $E_p$ in the $\nu^{-\beta}$ part of the SED is then 
\begin{equation}
F_x \propto (t-t_p)^{-2-\beta-a-b\beta}\,,
\end{equation}
while the LAE optical light curve (below the peak energy, in the 
$\nu^{1/3}$ part of the SED) is described by
\begin{equation}
F_{\rm opt} \propto (t-t_p)^{-5/3-a+b/3}\,.
\label{nonLAE}
\end{equation}
To check this model, we fixed the peak energy to $E_p= 4.3$ keV at
epoch 1 and the spectral slope to $\beta=0.72$ (Table \ref{tab:joint}). We fitted  the X-ray and optical
data between 205 and 250~s after the trigger,  i.e., between epochs 1 and 3,
when the optical light curve was rising. This gives $t_p =
185.9\pm7.5$ ~s, $a=-1.7\pm0.2$, and $b=1.7\pm0.5$, where $a$ and $b$ 
follow  from the derived decay slopes via Eq.~\ref{nonLAE}. 
Figure \ref{fig:peaksed2} shows how the fit is able to follow the SED
during epochs 2 and 3. 
The fit puts the time zero-point at the beginning of the main emission of the
proper GRB. This finding is qualitatively  in line
with other studies of other X-ray afterglows (e.g. \citealt{Liang2006}).

While the fit is satisfactory, one might wonder why at epoch 1 the
low-energy part of the SED touches the optical data point only within
1~$\sigma$.  However, there is actually
much more uncertainty in the extinction-corrected  UVOT flux than is given simply 
by the measurement error of 0.25 mag (\emph{white} filter, 
\citealt{Roming2009a}; see
Table~\ref{tab:magsUVOT}). The biggest uncertainty\footnote{A 
smaller uncertainty comes 
from the Galactic reddening derived from \citet{Schlegel1998}, 
which percentage error can be large for low reddening values.}
comes  from the correction for extinction in the GRB host
galaxy. Assuming a Milky  Way extinction law, a ratio of total-to-selective
extinction of $R_V$=3.08 (i.e., the standard value), and $A_V^{\rm host}=0.12$
mag (Table~\ref{tab:AV})  gives  a correction for host extinction for the UVOT
white filter of 0.52 mag (including the cosmological $k$-correction and the
correct CCD sensitivity characteristics for UVOT/white filter
observations\footnote{http://heasarc.gsfc.nasa.gov/docs/heasarc/caldb/data/swift/uvota/}).
  However,
$R_V$ in the  star-forming region where the GRB went off is not known exactly.
 Its $1\sigma$  error might well be on the order of 50\%. Finally, the
host extinction we have  derived here (Table~\ref{tab:AV})  is based on data
taken 20~ks after the burst.  It is an open question whether the host extinction
was already the same amount 200~s after the onset of the burst. In other
words, that the UVOT  white filter measurement does not exactly correspond
to the low-energy SED extrapolated from the X-ray data should not be
overinterpreted. However, it  naturally affects our test of the LAE model
since it introduces additional uncertainties.

\subsection{The afterglow phase}
\subsubsection{The light curve \label{betaXX}}

At early times, up to 470 s after the trigger, the X-ray light  curve is
dominated by two strong peaks (Fig.~\ref{fig:lcspec}).  The first peak is 4
seconds after the peak seen by BAT and GBM.  The optical light curve is
similarily complex, showing bumps up to about 10 ks after the
trigger. Unfortunately, the gap in the X-ray data does not allow a
comparison between the two bands during this timespan.

Despite the rich variability in the early afterglow, the late-time
evolution is consistent with a power law decay.
After 4.2~ks, the X-ray light curve can be described by a  broken power law
(\citealt{Beuermann1999a}) with  $\alpha_1^X =0.72 \pm 0.35, \alpha_2^X = 1.87
\pm 0.07, t_{\rm b} = (8100 \pm 1600) \rm{s}$ (observer frame) and a
fixed smoothness parameter $n=5 \ (\chi^2/\rm{d.o.f.} = 55.4/33 = 1.68$;
Fig.~\ref{fig:lcspec}). The optical data do not allow for a fit with a broken
power law.  For $t_{\rm obs}> 10$~ks the fit with a single power law gives
$\alpha^{\rm opt}  = 2.17 \pm 0.02\ (\chi^2/\rm{d.o.f.} = 56.8/34= 1.67)$.
The optical/NIR and X-ray data suggest similar small variability 
after 20~ks, which however we cannot study further for lack of good data. 
The break in the X-ray light curve could be a jet break,
but as we argue later, our detailed modeling of the afterglow
does not support this conclusion (Sect.~\ref{sec:inject}).

\begin{figure}[t]
\includegraphics[width=8.9cm,angle=0]{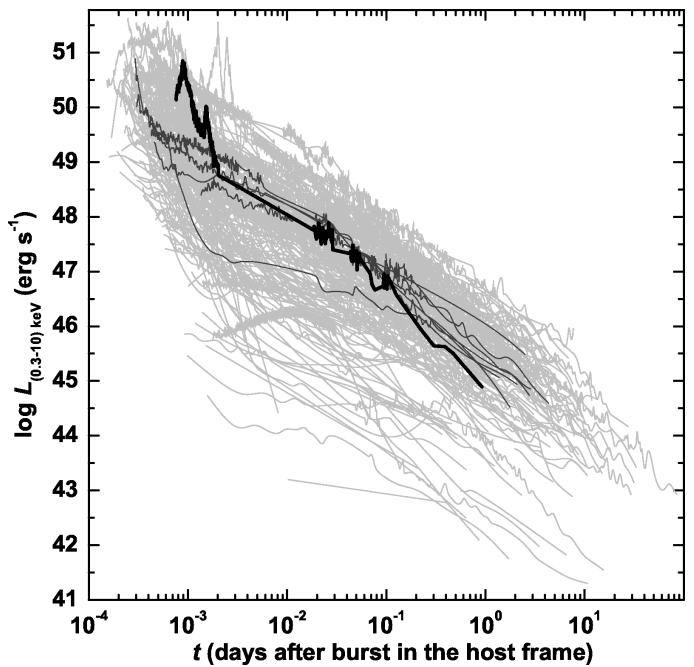}
\caption{
The X-ray luminosity of 190 \swift~GRBs and their afterglows  in the range of
0.3 to 10 keV between Jan 26, 2005, and Apr 25, 2010. \object{GRB 080928} is shown
in black. For comparison all six GRBs within a redshift interval of 0.1 around
the redshift of \object{GRB 080928} are highlighted in dark gray. The luminosity of the
afterglow of \object{GRB 080928} was basically in the mean of the X-ray luminosities
that have so far been observed.}
\label{fig:xlc}
\end{figure}

In Fig.~\ref{fig:xlc} we compare the X-ray afterglow of \object{GRB 080928} with all
X-ray afterglows found up to April 2010  in a redshift interval of $\Delta
z=0.1$ around the redshift of \object{GRB 080928} ($1.6919$), namely \object{GRB 050802} ($z =
1.7102$; \citealt{Fynbo2009a}), 071003 ($z=1.60435$; \citealt{Perley2008b}),
080603A ($z=1.6880$; \citealt{Perley2008a}), 080605 ($z = 1.6403$;
\citealt{Fynbo2009a}), 090418 ($z=1.608$; \citealt{Chornock2009a}), 091020 ($z
= 1.71$; \citealt{Xu2009a}), and  100425A ($z=1.755$; \citealt{Goldoni2010}).
In comparison to these, the early X-ray emission of \object{GRB 080928} is about 1.6
dex more luminous, probably  thanks to its physical connection to the prompt
emission.  Even compared to the entire ensemble of 190 X-ray light curves, it
is more luminous than the average.  However, after the light curve  break at
8.1~ks (observer frame; 3~ks host frame), the afterglow  rapidly becomes
subluminous with respect to the ensemble. Interestingly, except for GRB
080603A and 100425A, the other afterglows have a similar break time and
post-break decay slope.

\begin{figure}[t]
\includegraphics[width=8.9cm]{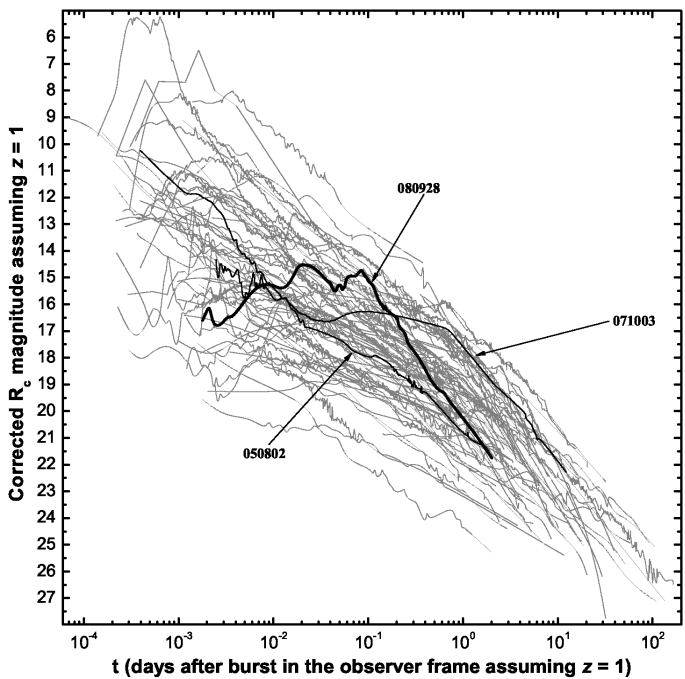}
\caption{
The optical afterglow of \object{GRB 080928} (thick line) compared  with the sample of
extinction-corrected afterglows shifted to $z=1$ from \cite{Kann2010}.
For comparison, the GRBs within a redshift interval of 
0.1 around the redshift of \object{GRB 080928} for which we have optical data
are highlighted and labeled. All magnitudes are Vega magnitudes.}
\label{fig:BigFig}
\end{figure}

In the optical bands the afterglow tends to vary between two extremes.  We
correct the afterglow for the extinction derived below (Sect.~\ref{betaOO})
and shift it to $z=1$ following \cite{Kann2006a}.  Compared to the ensemble of
optical afterglows with reasonable data \citep{Kann2010}, at early times it is
comparatively faint, nearly eight magnitudes fainter than the brightest events
(Fig.~\ref{fig:BigFig}).  Its multiple rebrightenings, which are a notable
signature of this afterglow, then bring the late-time light curve close to the
mean magnitude of the distribution at one day after the GRB (at $z=1$).  In
between, at about 0.1 days (at $z$=1), they make the afterglow about 2 mags
brighter than the average, shifting it into the group of the ten top brightest
optical afterglows at that time.

\subsubsection{The broad-band SED}
\label{betaOO}

\begin{figure*}[thb]
\includegraphics[width=18.4cm]{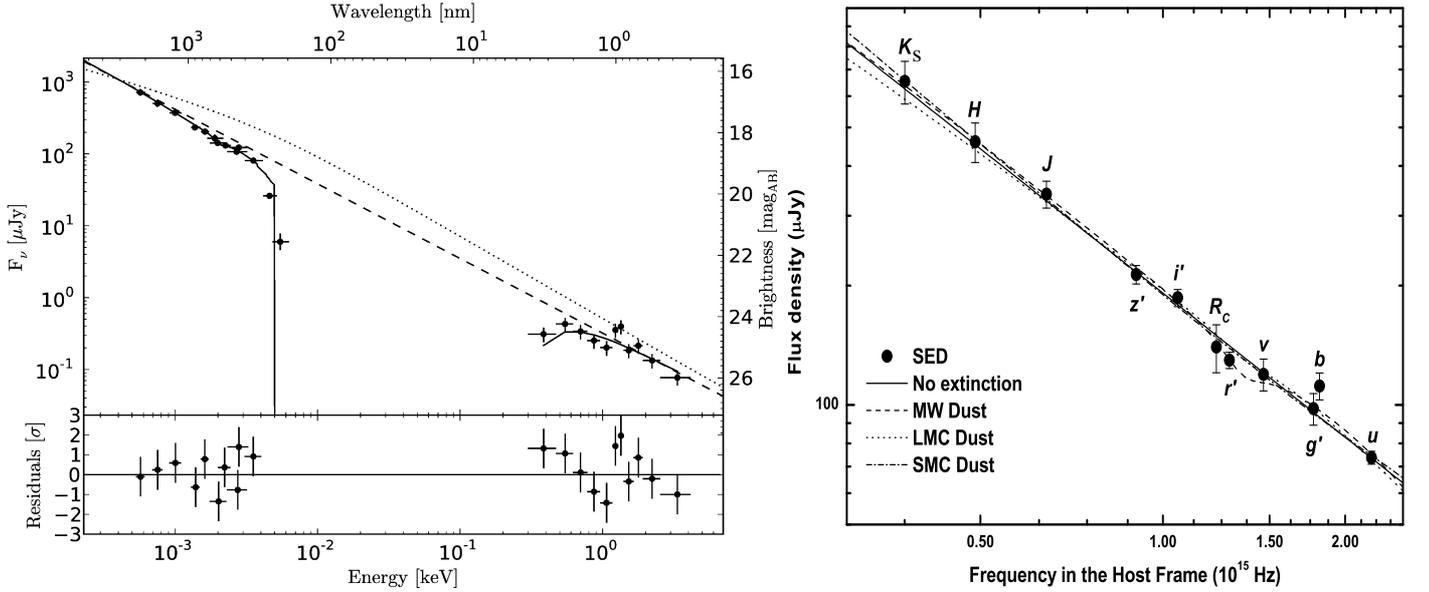}
\caption{
The observed SED of the X-ray/optical/NIR afterglow of \object{GRB 080928} at $t$=20 ks
after correction for Galactic extinction by dust (Table~\ref{SED})
and Galactic absorption by
gas. {\it Left:} The joint X-ray/optical SED is almost a pure power law
(dashed line) affected by only a small amount of host extinction by dust
(Table~\ref{tab:AV}) and $3.5\,\times\,10^{21}$ cm$^{-2}$ of host absorption
by the gas. The UV bands are affected by Lyman drop-out. The dotted line
represents the SED that follows from the numerical energy injection model for
this particular time, which  slightly overpredicts the flux in the X-ray band
(see Sect.~\ref{sec:inject}). Residuals refer to the  the plot with
$\beta_{\rm OX}$ = 1.02 (broken line).  {\it Right:} Zoom-in to the
optical/NIR SED, and the different dust models used to fit the data,  where it
is possible to discern the dip resulting from the 2175~{\AA} feature.}
\label{fig:grbsed}
\end{figure*}

To fit the unabsorbed SED from the optical to the X-ray bands, we
selected the X-ray data from 12.4 ks to 25 ks (mean photon arrival time 20
ks). Since no evidence of any color variations was found in the optical data, we
then shifted the optical light curve to this time (Table~\ref{SED}; corrected
for a Galactic extinction of $E(B-V)=0.07$ mag). In addition to the 
GROND  and UVOT data we used the VLT detection corrected to the $R_C$ band 
 (Sect.~\ref{sec:optnir}). In doing the fit, we fixed the redshift to
1.69, the host galaxy hydrogen column density to $N_{\rm H}=
3.5\,\times\,10^{21}$ cm$^{-2}$ and the Galactic hydrogen column density to
$N_{\rm H}=0.56\,\times\,10^{21}$ cm$^{-2}$ (Sect.~\ref{XAG}). The resulting SED is
shown in Fig.~\ref{fig:grbsed} (left) and Table~\ref{tab:AV}.  There is no
spectral break between the X-ray band and the optical.  Between 4 ks until the
end of the X-ray observations at around 120 ks (1.4 days) no evidence of
spectral evolution was found.

\begin{table}[t]
\begin{minipage}{\columnwidth}
\renewcommand{\footnoterule}{}
\centering
\caption{The values plotted in Fig.~\ref{fig:grbsed}, corrected for 
Galactic extinction and given in Vega magnitudes, obtained at $t=20$~ks.}
\label{SED}
\begin{tabular}{lrrcc}
\toprule
Filter & $\lambda$ \hspace*{3mm} &  $\nu$(1+z) &   mag &  $F_\nu$\\
       & (nm)   \hspace*{1mm}  &($10^{14}$Hz)&      & ($\mu$Jy)\\
\midrule
$K_S$  & 2151.2  &   3.75  &  $15.07  \pm 0.13$    & $653.8 \pm102.4$\\
$H$    & 1646.7  &   4.90  &  $15.86  \pm 0.12$    & $460.7 \pm52.7 $\\
$J$    & 1256.1  &   6.42  &  $16.66  \pm 0.09$    & $340.0 \pm26.6 $\\
$z'$   & 893.0   &   9.04  &  $17.56  \pm 0.06$    & $213.4 \pm11.4 $\\
$i'$   & 762.6   &   10.58 &  $17.84  \pm 0.05$    & $186.2 \pm 9.29$\\
$R_C^*$  & 658.8   &   12.25 &  $18.36  \pm 0.15$    & $140.0 \pm19.4 $\\
$r'$   & 627.0   &   12.87 &  $18.47  \pm 0.05$    & $129.6 \pm6.05 $\\
$v$    & 550.5   &   14.66 &  $18.71  \pm 0.10$    & $119.4 \pm11.0 $\\
$g'$   & 455.2   &   17.73 &  $18.91  \pm 0.10$    & $97.8  \pm9.07 $\\
$b$    & 444.8   &   18.14 &  $18.90  \pm 0.08$    & $111.7 \pm8.64 $\\
$u$    & 365.2   &   22.10 &  $18.48  \pm 0.04$    & $73.7  \pm2.81 $\\
$uvw1$ & 263.4   &   30.64 &  $18.92  \pm 0.11$    & $23.8  \pm2.38 $\\
$uvm2$ & 223.1   &   36.17 &  $20.53  \pm 0.30$    & $5.40  \pm1.51 $\\
$uvw2$ & 203.0   &   39.76 &  $>19.81         $    & $<11.4         $\\
\bottomrule
\end{tabular}
\tablefoot{
The $R_C$-band value is based on \cite{Vreeswijk2008a}; the other data
refer to the GROND and the UVOT bands.}
\end{minipage}
\end{table}

\begin{table}[t]
\begin{minipage}{\columnwidth}
\renewcommand{\footnoterule}{}
\centering
\caption{Results of the joint optical to X-ray spectral fit.}
\begin{tabular}{cccl}
\toprule
Dust model & $A_V^{\rm host}$ & $\beta_{\rm OX}$ & $\chi^2/$d.o.f \\
\midrule	
 MW & $0.12\pm0.03$ & $1.03 \pm 0.01$ & 20.2/18 \\
LMC & $0.07\pm0.02$ & $1.02 \pm 0.01$ & 24.5/18 \\
SMC & $0.04\pm0.01$ & $1.01 \pm 0.01$ & 26.6/18 \\
\bottomrule
\end{tabular}
\tablefoot{Columns 2,3:
$A_V^{\rm host}$ is the deduced host extinction and $\beta_{\rm OX}$
is the optical to X-rays spectral slope.}
\label{tab:AV}
\end{minipage}
\end{table}

We find that SMC and LMC dust provided an acceptable fit, although Milky Way (MW) dust
improved the fit (Table~\ref{tab:AV}). The 2175\AA\ feature is  weaker
than in the case of \object{GRB 070802}  \citep{Kruhler2008a, Eliasdottir2009},
however.  The derived host extinction is  clearly unremarkable within the
sample of \citet{Kann2010}.

For a MW interstellar medium the deduced high $N_{\rm H}$ would  imply
a host extinction of $A_V^{\rm host}= 2^{+1.0}_{-1.2}$ mag, in contrast to 
the low value found here. However, several GRB afterglows studies have 
found that, despite a very large scatter in the $N_{\rm H}/A_V$ ratio, the $N_{\rm H}$
is always significantly greater than observed in the local Universe (e.g.,
\citealt{Galama2001,Stratta2004,Kann2006a,Starling2007,Schady2007,Schady2010}),
a phenomenon that could potentially be explained by dust destruction 
by the intense fireball light (\citealt{Fruchter2001,Watson2007}).

\subsubsection{Theoretical modeling of the light curve \label{sec:inject}}

Using the forward shock afterglow model \citep[e.g.,][]{Panaitescu2000,Zhang2004a,Piran2005},
 it is difficult to explain
the different slopes of the optical and X-ray light curves given that they are
on the same power law segment of the spectrum.  Assuming the cooling
frequency, $\nu_c$, is above the X-ray band, the spectral slope gives an
electron energy index of $p = 2\beta + 1 \approx 3$.  The light curve slope of
$\alpha \approx 2$ then indicates we have a pre-break evolution in a stellar
wind.  This would be problematic for the early-time evolution, as it is
difficult to get a rising afterglow with a stellar-wind external medium.  The
second possibility is that $\nu_c$ is below the optical bands, resulting in $p
= 2\beta \approx 2$.  The light curve slope then indicates we are in a
post-break evolution.  If the external medium is constant, then this does not 
contradict the early-time observations, given a small enough initial
Lorentz factor. Having $\nu_c$ below the optical bands is, however, difficult
to achieve as we show below.

The early optical light curve is rich in variability. Unfortunately, there are
no XRT measurements during the optical fluctuations to verify the correlation
between X-ray and optical light curves, but there are a couple of other cases
where high-energy flares are seen in the optical, too, e.g., \object{GRB 041219A}
\citep{Vestrand2005, Blake2005}, \object{GRB 050820A} \citep{Vestrand2006}, \object{GRB 060526}
\citep{Thoene2010}, \object{GRB 061121} \citep{Page2007a}, and \object{XRF 071031}
\citep{Kruhler2009a}.  In particular, the general behavior of the afterglow
recalls the cases of  \object{GRB 060904B} \citep{klotz2008a, Kann2010} and \object{GRB 060906}
\citep{Cenko2009a}. The optical fluctuations have a long timescale that is
more consistent with energy injection into the forward shock than with central
engine activity.

To fit the afterglow data we used the numerical model of  \citet{Johannesson2006a} and
\citet{Johannesson2006b}, with modifications as described in \citet{Perez-Ramirez2010}.
We excluded data taken
in the first $500$~s after the trigger, as they are most likely explained by
internal shocks.  The data are still kept in the fit as upper limits: not
considered if the model is below them, but added to the $\chi^2$ value like normal
points if the model is above them. We explored two different times as the initial
time for the calculation: the trigger time $t_0$ and the start of the main
prompt emission at $t_0 + 170\,$s.  Since a wind-like medium will overpredict
the early data, we limited our study to a constant-density medium.  Our
assumptions were that the first peak in the optical light curve at $\sim$1000~s
is the onset of the afterglow and that the following two bumps at $\sim$2~ks and
$\sim$10~ks are caused by energy-injections.  Host extinction was assumed to
be due to Milky Way dust (Table~\ref{tab:AV}), but we allowed
$A_V^{\rm host}$ to be free during the fit.  We accounted for Ly~$\alpha$
extinction with the method of \citet{Madau1995}.

In the forward shock model it is generally assumed that the shock front
expands sideways at the speed of sound. Using numerical calculations,
\citet{KumarGranot2003} find that the expansion speed of the jet is
significantly lower than this simple estimate.  One of the effects of a slower 
sideways expansion is that the jet break is reached later in the evolution.
This poses some problems when fitting the sharp overturn after the last
optical bump, because the energy-injections effectively move the evolution of the
forward shock back in time.  We have found that reducing the expansion speed
to $\sim$20\% of the speed of sound mitigates this problem, in agreement with
the values found by \citet{KumarGranot2003}.  We note that this is an upper
limit on the expansion speed, since lower values can be used to explain the data.

\begin{figure}[htb]
\includegraphics[width=8.9cm]{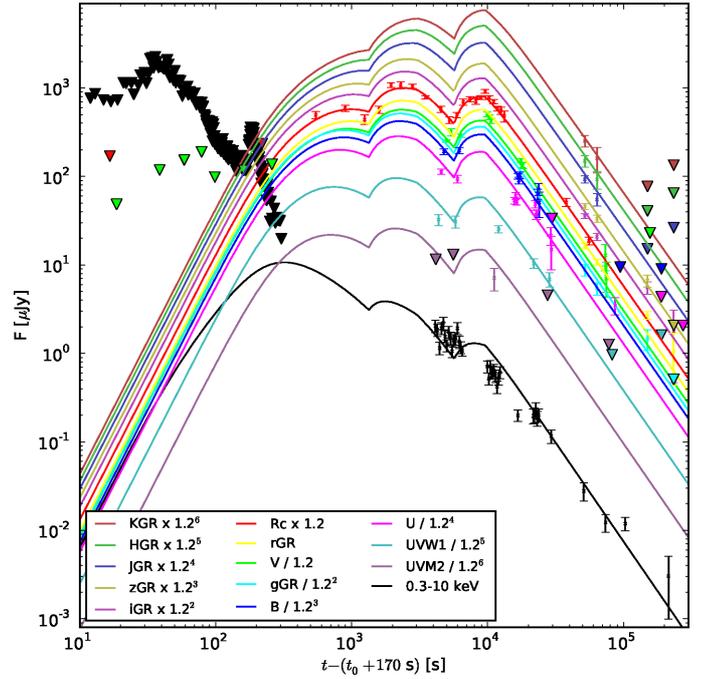}
\caption{The best-fit light curves of the afterglow of \object{GRB 080928}. 
The agreement between the model and the observational data is best
if the reference time is shifted by 170 seconds.
The parameters of the model are given in Table~\ref{tab:tabinject}.
Filters called ``GR'' stand for the GROND filter set. Light curves
in different bands are arbitrarily shifted for clarity by powers of 1.2.}
\label{fig:lc_Gulli}
\end{figure}

Table~\ref{tab:tabinject} gives the parameters of the best-fit model shown in
Fig.~\ref{fig:lc_Gulli}.  The numerical model prefers the start time
of $t_0+170\,$s where most of the constraints come from the optical data
contemporaneous with the high-energy prompt emission.  The model overpredicts
the data in this epoch when the start time is $t_0$.  The best fit results in
$\chi^2/\rm{d.o.f.} = 307/187 = 1.64$, which is comparable to the power law
fits shown earlier despite fitting more data.  We note that the fit does not
do a good job with the X-ray light curve, slightly underpredicting it before
the second injection and then overpredicting it afterwards.  This seems to
indicate that there is some other mechanism at work than energy-injections,
but the lack of simultaneous X-ray observations during the optical rise makes
it difficult to say what is going on.

Unfortunately, we are unable to find a suitable set of initial
parameters such that we have $\nu_c$ below the optical frequency and do
not overpredict the flux.  This is caused by the fact that in post-break
evolution we have \citep{Rhoads1999}
\begin{eqnarray}
	\nu_c &\propto& \epsilon_B^{-3/2}n_0^{-5/6}E_0^{-2/3}, \\
	F_{\rm max} &\propto& \epsilon_B^{1/2}n_0^{1/6}E_0^{4/3},
\end{eqnarray}
where $F_{\rm max}$ is the afterglow flux at the peak frequency, $\epsilon_B$ 
the fraction of energy contained in the magnetic
field, $n_0$ the density of the external medium, and $E_0$ the
initial energy release.  As we see from these equations, it is
very difficult to lower the value of $\nu_c$ without increasing the flux of
the afterglow. This can be overcome by placing
the break frequency close to the optical waveband and increasing the
absorption.  The spectrum from the numerical fit is shown in
Fig.~\ref{fig:grbsed}, and it explains the data equally well as a single 
power law.  One must also note that the cooling
break is not sharp, because we are integrating over the equal arrival time
surface with different intrinsic values for the cooling break.

The error estimates given in Table~\ref{tab:tabinject} are found from a
$\chi^2$ profile method, and we consider these errors  to be reliable.  Due to
lack of radio and mm data, our limit on $n_0$ is mostly from the requirement
for an early jet-break, although the low value of $\nu_c$ also plays a role.
The limit on the initial Lorentz factor, $\Gamma_0$, is found from the
requirement that the first optical bump coincides with the onset of the
afterglow.  The low value of $\Gamma_0 \sim 100$ favors a high-energy
spectral slope of 2.5  as indicated by the Band function fit in
Table~\ref{tab:gbm_spec_one}. 

The initial half opening angle of the jet, $\Theta_{\rm 0}$, has an
unusually low value, required by the assumed small jet break time of 10 ks.
This low value is also needed to model the rapid change in the light curve
slope during the energy injection episodes.  The shape of the light curve
after energy-injections is determined by the relativistic aberration of the
forward shock light and therefore $\Theta_{\rm 0}$.  We note that this low
value depends on the assumed geometry of the forward shock, here assumed to
be isotropic and spherical within the narrow confinement region.

The large energy-injections are actually a feature of the energy-injection 
model, and these values are compatible with other studies using this model 
(\citealt{Thoene2010}, \citealt{deUdarte2005a}).  For the energy 
injected to have a visible effect on the light curve, the energy has to 
be compatible with the energy in the shock front, leading to an ever 
increasing energy of the injections.  We also note that the total energy 
budget of the afterglow is highly uncertain, mostly caused by the large 
uncertainties in the values of $\epsilon_B$ and $\epsilon_i$ that require 
broader energy coverage in the data to be properly constrained.
Limits on other parameters are found from the general
spectral and light-curve evolution of the afterglow and are more robust
against the assumed start time.

\begin{table}[htb]
\begin{minipage}[t]{\columnwidth}
\renewcommand{\footnoterule}{}
\centering
\caption{Parameters deduced for the energy-injection model.}
\label{tab:tabinject}
\begin{tabular}{lcr}
\toprule
\multicolumn{3}{c}{Parameters of the energy-injection model} \\
\midrule
$E_{\rm total}$ & $1.4_{-0.7}^{+45}\,\times\,10^{50}$ erg & total released energy \\[0.7ex]
$E_0$          & $1.5_{-1.0}^{+28}\,\times\,10^{49}$ erg  &initially released energy\\[0.7ex]
$E_1$          & $2.0_{-0.9}^{+2.3}\,\times\,E_0$  & energy of the first injection \\[0.7ex]
$t_1$          & $22_{-3}^{+2}$ minutes &time of the first injection \\[0.7ex]
$E_2$          & $5.7_{-1.8}^{+4.4}\,\times\,E_0$ &energy of the second injection \\[0.7ex]
$t_2$          & $95_{-4}^{+3}$ minutes &time of the second injection \\[0.7ex]
$\Gamma_0$     & $77_{-28}^{+120}$ &initial outflow Lorentz factor  \\[0.7ex]
$n_0$          & $29_{-28}^{+500}$ cm$^{-3}$ & circumburst medium density\\[0.7ex]
$\Theta_{\rm 0}$& $0.50_{-0.22}^{+0.65}\deg$      & initial half-opening angle \\[0.7ex]
$p$            & $2.29_{-0.11}^{+0.06}$    & electron index\\[0.7ex]
$\epsilon_i$   & $0.037_{-0.035}^{+0.060}$   & fraction of energy in the\\
               &                            & lowest-energy electrons\\[0.7ex]
$\epsilon_B$   & $2.5_{-2.4}^{+16}\,\times\,10^{-4}$ &magnetic energy fraction\\[0.7ex]
$A_V^{\rm host}$& $0.37_{-0.09}^{+0.07}$ & host (MW) extinction\\
\bottomrule
\end{tabular}
\tablefoot{All times given are in the observer frame
relative to the start time of $t_0 + 170$~s. See Sect.~\ref{sec:inject}.}
\end{minipage}
\end{table}

Do the parameters obtained from the modeling of the afterglow light curve 
agree with the LAE model (Sect.~\ref{sec:largeangle})? 
If the the X-ray tail is LAE, then the observer
time is the photon arrival time from a region moving at an angle $\Theta$, so
that
\begin{equation}
t-t_p = (1+z)\,(R/c)\ [1/(2\Gamma^2)\, + \,\Theta^2/2]\,,
\end{equation} 
where $\Gamma$ is the Lorentz factor of the outflow and $t_p$ is the
zero-point of the beginning of the main emission of the proper GRB.  The peak
time of the GRB, $t_e$, corresponds to the arrival-time of  photons emitted
from an angle $\Theta=1/\Gamma$, which implies that 
$t_e-t_p = (1+z)\,(R/c)/\Gamma^2$, and hence
\begin{equation}
(t-t_p)/(t_e-t_p)= [1 + (\Gamma\, \Theta)^2] / 2\,. 
\end{equation}
Since the outflow has a finite half-opening angle, $\Theta_{\rm 0}$,
the LAE can be seen only up to a time $t_{\rm max}$ given by
\begin{equation}
t_{\rm max}-t_p = (t_e-t_p) \ [1 + (\Gamma \, \Theta_{\rm 0})^2]/2\,.
\label{eq:tmaxlae}
\end{equation}
In section~\ref{sec:largeangle} we found $t_p=185.9\pm7.5$~s, and we argued
that the LAE emission should have been active at least until  the third
optical observing epoch, which sets  $t_{\rm max}>250$~s. In addition we found
that the proper burst has its main peak at $t_e=204$~s. For these numbers
eq.~(\ref{eq:tmaxlae}) gives $\Gamma \, \Theta_{\rm 0} \gtrsim 2.4$, a
relation that is fulfilled by the model within the errors
(Table~\ref{tab:tabinject}). 

Finally, could there be a possible contribution from a reverse shock?
Basically, there is only one observational constraint (the optical flux)
among many parameters that determine the reverse shock emission. Given
that our model requires a substantial energy injection in the forward shock,
there could be a substantial optical emission from a long-lived reverse
shock, so there could be a significant contribution to the optical bumps
from the reverse shock. In the numerical model we have only considered the
forward shock because that shock is more likely to be 
the source of the X-ray emission after 10 ks, i.e., when
energy injection ceases and the ejecta electrons cool fast enough to yield
little X-ray emission. Adding the contribution of the reverse shock(s) to
the model might not affect the value we obtained for the jet opening
angle.

\subsection{The isotropic equivalent energy and gamma-ray peak luminosity
\label{isoequi} }

Given the results of the spectral fit in the high-energy domain, we can
estimate the isotropic-equivalent energy released during the prompt emission
phase. Fitting the BAT and GBM data for the time of the
gamma-ray precursor between 46.5~s and 121~s  gives an isotropic equivalent
energy of $E_{\rm iso}$ (1-10000 keV) = (0.40$\pm0.03) \,\times\, 10^{52}$
erg, while a fit of the  combined XRT-BAT-GBM data during the main peak
emission between $t_0+198.75$~s and $t_0+228.4$~s leads to $E_{\rm iso} =
(0.88\pm0.025) \,\times \,10^{52}$ erg.  Fixing the peak energy for
the value found in the second interval ($132_{-16}^{+49}$ keV; 
Table~\ref{tab:gbm_spec_one}), we find for the
whole burst from $t_0-23.5$~s to  $t_0+372.5$~s an isotropic energy of $E_{\rm iso}=
(1.44\pm0.92)\,\times\,10^{52}$ erg, in agreement  with the Amati relation
\citep{Amati2006a}.

From the light-curve modeling in Sect.~\ref{sec:inject} we obtained 
$\Theta_{\rm 0}$ and $E_0$ (the energy in
the collimated ejecta;
Table~\ref{tab:tabinject}), so that the 
isotropic equivalent kinetic energy $E_{\rm kin, iso}$ can be calculated. 
This energy, when compared to $E_{\rm iso}$,
gives the radiative efficiency $\eta$ in the prompt emission phase,
$\eta = E_{\rm iso}/(E_{\rm kin, iso}+E_{\rm iso})$.
Unfortunately, within the $1~\sigma$ error bars of the model fit the result 
is not constraining.

The \fermi/GBM data allows an estimate of the variability of the light
curve, a quantity that has been shown to correlate with the isotropic
equivalent peak luminosity, $L_{\rm iso, peak}$. Following the method described
in \citet{li2006} (see also \citealt{Rizutto2007}) 
and using a smoothing time scale of $t_{50}=3.3$\,s, we
derived a variability index of $V=-2.67$, which is the normalized squared
deviation of the observer-frame light curve from a Savitzky-Golay filtered
reference light curve. This results in log $L_{\rm iso, peak}$ [erg
s$^{-1}$]=$50.75^{+0.49}_{-0.59}$ (100\,keV to 1\,MeV, rest frame),
about three orders of magnitude less than in the case of 
the very energetic burst \object{GRB 080916C} \citep{Greiner2009a}.

\begin{figure*}[htb]
\centering
\includegraphics[width=18.4cm,angle=0]{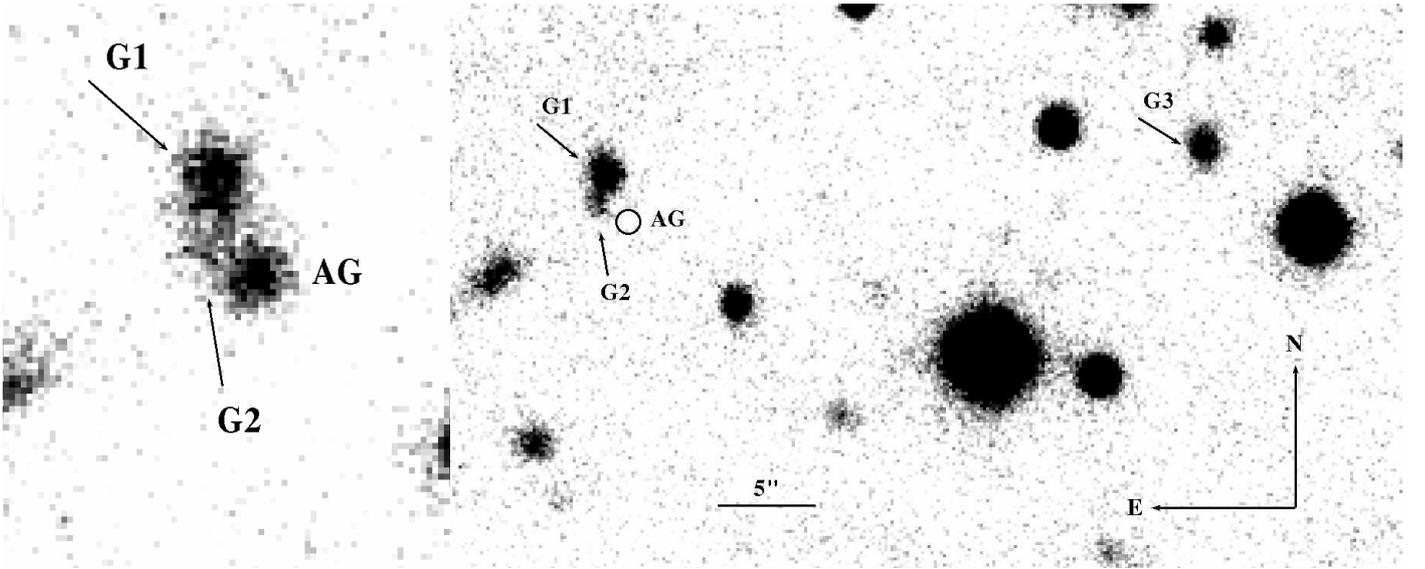}
\caption{
{\it Left:} 
Zoom-in of the GROND combined $g^\prime r^\prime i^\prime z^\prime $-band
image obtained 1.74 days after the burst at a seeing of 1\farcs5. It shows the
afterglow (AG) and the brightest galaxies close to it. {\it Right:} Zoom-in of
the stacked GROND optical $g^\prime r^\prime i^\prime z^\prime $-band images
obtained on May 15, 2009, 6.5 months after the burst (5th epoch) when the
afterglow had faded away. It also shows the galaxy (G3) that was
coincidentally covered by the slit of the spectrograph 
when the redshift of the afterglow
was measured with the  ESO/VLT \citep{Vreeswijk2008a, Fynbo2009a}. Data for G1
to G3 are summarized in Table~\ref{tab:g1-3}.}
\label{fig:epozoom}
\end{figure*}

\subsection{The GRB host galaxy \label{hosts}}

The deep fifth-epoch GROND images taken 6.5 months after the burst at a seeing
of $\sim 1''$ do not show any galaxy underlying the position of  the optical
transient down to the following 3$\sigma$ upper limits (AB magnitudes):
$g^\prime =25.4, r^\prime =25.6, i^\prime =24.6, z^\prime =24.3, J=22.0,
H=21.6, K_S=20.9$. Assuming for simplicity a power law spectrum for this
galaxy of the form $F_\nu \propto \nu^{-\beta_{\rm gal}}$, for the $r^\prime $
band this translates into an absolute magnitude of  $M_{r^\prime }=
m_{r'\prime} - \mu - k$, where $\mu=45.54$ mag is the distance modulus and $k$
 the cosmological $k$-correction, $k= -2.5 (1-\beta_{\rm gal}) \log
(1+z$). For a representative value of $\beta_{\rm gal}$=1, this gives a lower
limit of $M_{r^\prime }> -19.94$, which agrees with the luminosities
found so far for the GRB host galaxy population.  In fact, 
much less luminous GRB hosts 
are known (see \citealt{Savaglio2009}). However, could one of the
galaxies seen in projection close to the afterglow be the host?

Close to the position of the afterglow there is a relatively bright galaxy
(labeled G1 in Fig.~\ref{fig:epozoom})  with $r^\prime=23.41\pm0.05$.  Using
the stacked GROND $g^\prime r^\prime i^\prime z^\prime $-band images from the
fifth epoch,  its central coordinates (Table~\ref{tab:g1-3}) are offset by
$2.6\pm0.3$ arcse from the position of the optical afterglow.  If this
galaxy is at the redshift of the burst, then  the projected offset of the
optical transient from its center is $22.2\pm2.6$ kpc. This is almost 20 times
more than the median projected angular offset of 1.31 kpc found by
\citet{Bloom2002a} for a sample of 20 host galaxies of long bursts, making it
unlikely that this is the host galaxy of \object{GRB 080928}.

Some arcseconds south of G1 lies a diffuse object that could either be
physically associated to G1 or represent another foreground/background
galaxy. This object (G2 in Fig.~\ref{fig:epozoom}) is $1.5\pm0.3$ arcsec
away from the afterglow position. If it is at the redshift of the burst, its
projected distance from the afterglow is $13\pm2.6$ kpc, again hardly in
agreement with the observed GRB offset distribution. However, both
objects/galaxies are potentially close enough in projection to imprint a
signal on the GRB afterglow spectrum. Indeed, \cite{Fynbo2009a} report a
foreground absorption line system exhibiting several strong Fe, Mg and Ca
lines  at a redshift of $z=0.7359$. In the $1\arcsec$ slit passing over the
afterglow, \cite{Fynbo2009a} identify a galaxy $30\arcsec$ away from  the
afterglow at a redshift of $z=0.736$. This redshift is identical to the value
found for the absorption line
system \citep{Vreeswijk2008a}. We labeled this galaxy as G3.

To clarify if G1 or G2 could be responsible for the absorption 
line system found in the afterglow light, we fit our multicolor
photometry of these galaxies using \emph{HyperZ} (\citealt{Bolzonella2000}).
This multicolor photometry was performed on the GROND
images in the following way. 
At first PSF-matching techniques under IRAF were used to 
correct for a different seeing (see \citealt{Alcock1999}).
Then aperture photometry was applied.
In Table~\ref{tab:hypresult} we provide the best fit 
of the observed broad-band SEDs of G1, G2, and G3 for 
a fixed redshift (either $z=0.736$, the redshift of the intervening system,
or $z=1.6919$, the redshift of the afterglow). 
The results are based on GROND data obtained 6.5 months after the
burst. They
indicate that with high probability none of the galaxies
is the host galaxy and that G1 is not the foreground absorber seen in the afterglow spectrum. 

For object G3, we find a \emph{HyperZ} solution in very good agreement with
the value of $z=0.736$ reported by \cite{Vreeswijk2008a} ($\chi^2$/d.o.f =
1.01). However, the detection in only the four optical bands does not allow us
to constrain the dust extinction in this galaxy.  Unfortunately, in the case
of G2 a \emph{HyperZ} fit with the redshift as a free parameter leads to no
conclusive results, the resulting error bars are very large, and the photometry
can be affected by the nearby galaxy G1. On the other hand, a  \emph{HyperZ}
fit with the redshift fixed at $z=0.736$ gives a reasonable photometric
solution  ($\chi^2$/d.o.f = 1.07; Table~\ref{tab:hypresult}). This makes it
possible that G2 is responsible for the absorption line system seen in the
afterglow spectrum, given the proximity of G2 to the  spectral slit passing
over G3 and the afterglow.

\begin{figure}[htb]
\includegraphics[width=8.8cm]{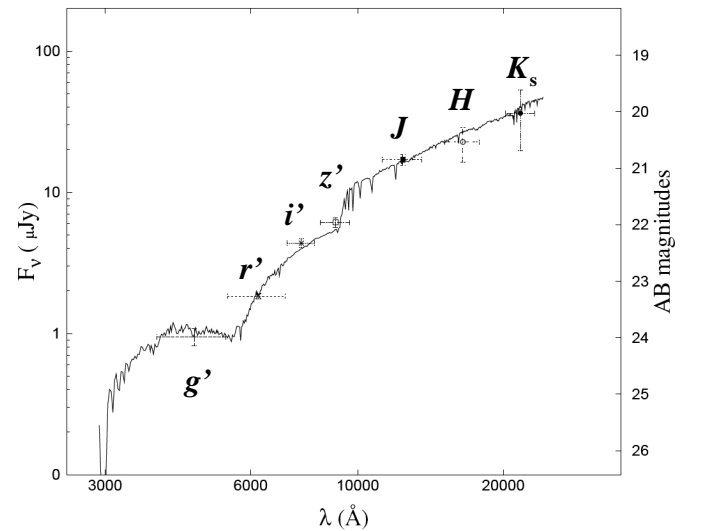}
\caption{
The SED of galaxy G1 close to the afterglow (see Fig.~\ref{fig:epozoom}),
obtained from images taken with GROND 6.5 months after the burst
($g^\prime r^\prime i^\prime z^\prime JHK_s$ filters).  Shown is the best 
\emph{HyperZ} fit  that is based
on the template of a dusty starburst galaxy at a redshift of $z$=1.46 (see
also Table~\ref{tab:hypresult}).}
\label{fig:galsed}
\end{figure}

When we treat the redshift as a free parameter, not fixing it to the value of
the afterglow or the absorbing system, we find that the best \emph{HyperZ}
solution for G1 is $z=1.46^{+0.15}_{-0.10}$ (Fig.~\ref{fig:galsed}), in both
cases (whether we consider G2 to be a separate galaxy or not), confirming that
G1 is not related to any other object.  

\begin{table*}[htb]
\caption{
Coordinates and AB magnitudes of objects G1 to G3, not corrected
for Galactic extinction.}
\begin{minipage}[t]{\textwidth}
\renewcommand{\footnoterule}{}
\centering
\begin{tabular}{l|c|ccc cccc}
\toprule
Object & R.A., Dec. (J2000) &  $g^\prime$  &  $r^\prime$  &  $i^\prime$ &  $z^\prime$  &  $J$  & $H$  &  $K_s$ \\
\midrule
G1 &06:20:16.96, $-$55:11:56.6  & 24.22(15)    &23.41(05)    &22.43(08) &22.03(08)    &20.89(09)  &20.55(30)   &20.00(50)\\
G2 &06:20:16.99, $-$55:11:58.0  & 25.20(50)    &24.50(06)    &23.26(09) &22.70(05)    &21.50(20)  &21.20(30)   &$>20.7$ \\
G3 &06:20:13.35, $-$55:11:54.9  & 23.13(12)    &23.12(05)    &22.63(07) &22.20(05)    &$>22.0$   &$>21.6$    &$>20.9$ \\
\bottomrule
\end{tabular}
\label{tab:g1-3}
\end{minipage}
\end{table*}

\begin{table}[hbt]
\begin{minipage}[t]{\columnwidth}
\renewcommand{\footnoterule}{}
\centering
\caption{\emph{HyperZ} results for the fit of the SED of G1, G2, and G3.}
\begin{tabular}{l|ccc|ccc}
\toprule
Object	&$\chi^2_{0.7359}$	&Dust	&$A_V^{\rm host}$	&$\chi^2_{1.6919}$
&Dust	&$A_V^{\rm host}$\\
\midrule												
G1  starburst  &3.98   &LMC    &1.0    &3.47   &SMC    &0.8 \\
G2  irregular  &1.07   &LMC    &0.7    &3.49   &MW     &0.8 \\
G3  irregular  &1.01   &--     &0.0    &3.76   &--     &0.0 \\
\bottomrule
\end{tabular}
\tablefoot{ Column 1 provides the galaxy template that fit the data best. Columns 3 and 6 
contain information about the deduced
extinction law. Columns 4 and 7 give the corresponding global visual extinction.
For further details see Sect.~\ref{hosts}.}
\label{tab:hypresult}
\end{minipage}
\end{table}

\section{Summary}

\object{GRB 080928} was a long burst that lasted for about 400 seconds. It was detected by
\swift/BAT and \fermi/GBM and was followed up by \swift/XRT and \swift/UVOT.
Ground-based follow up observations were performed by 
the robotic ROTSE-IIIa telescope in Australia and the
multi-channel imager GROND on La Silla. Its early X-ray light curve is
dominated by two bright peaks that occurred within the first 400 seconds after
the BAT trigger. The first peak is delayed by some seconds from the
gamma-ray peak emission, while the second peak has no obvious counterpart in
the high-energy band. It occurred when the gamma-ray emission had already faded
away. After a data gap between about 400 s and 4 ks, the X-ray light curve
continued to show evidence of small-scale fluctuations, while between 200 s
and 10~ks the optical light curve shows bumps and dips, possibly related to
energy-injections into the forward shock (refreshed shocks).

Between about 200~s and 400~s after the BAT trigger, both \swift/UVOT and
 ROTSE-IIIa detected optical emission and \swift/XRT monitored X-ray 
radiation, while the GRB was still emitting in the gamma-ray band. 
The combination of these data allowed us to construct the SED from about 1 eV 
to 150 keV at several epochs, making \object{GRB 080928}
one of the rare cases where a spectral energy distribution spanning from
optical  to gamma rays can be traced during the prompt emission.  The first
epoch covers the  main peak emission in gamma rays, as well as in the X-ray
band. The resulting SED can be understood as due to synchrotron radiation
with a break energy around 4 keV. 

In addition, the optical and X-ray data allowed us to 
confirm that the radiation 
following the first strong peak seen in the X-ray light curve comes from
large-angle emission. The peak itself might have a different origin.
Considering the observed rising optical emission contemporaneous
to the decaying X-ray tail, we found that the data can only be understood 
if one of the assumptions made in the LAE model is relaxed, namely 
the assumption that the electron population is the same at all angles $\theta$.
This implies the use of a generalized version of the LAE model,
for which we obtain the flux and the energy of the peak evolving as 
$F_p \propto \theta^{-2-a}$ and $E_p \propto \theta^{-1-b}$, with
$a=-1.2\pm0.2$ and $b=1.1\pm0.5$. Those dependencies reflect the 
distribution with angle of the ejecta parameters that determine $F_p$ 
and $E_p$, such as ejecta kinetic energy per solid angle or the bulk 
Lorentz factor.

The X-ray data can be best fit by assuming  an effective hydrogen column density
in the host of $N_{\rm H}^{\rm host} = 3.6^{+1.8}_{-2.2} \times
10^{21}\,\rm{cm}^{-2}$. For a MW  interstellar medium, this would  imply
a host extinction of $A_V^{\rm host}= 2^{+1.0}_{-1.2}$ mag, in contrast to
$A_V^{\rm host}=0.12\pm0.03$ mag 
found in the optical afterglow data, which indeed
seem to favor a MW interstellar extinction law. That the
dust-to-gas ratio is relatively small along GRB sight-lines in their host
galaxies is a well-known phenomenon, possibly owing to dust destruction by the
intense fireball light.

In our interpretation of the data, the first peak in the optical light curve
at $\sim$1~ks is the onset of the afterglow, and the following two bumps at
$\sim$2~ks and $\sim$10~ks are caused by energy-injections. Applying an
energy injection model, the analysis explains the data after 10~ks  with a
post-jet evolution requiring a small opening angle ($\lesssim$~1.0 degree).

The optical afterglow was found to be about 2.6 arcsec south of a relatively
bright face-on galaxy, with unknown redshift. However, its photometric
redshift based on GROND $g^\prime r^\prime i^\prime z^\prime JHK_s$ data is in
disagreement with the redshift of the afterglow found by
\citet{Fynbo2009a}. In addition, the angular offset of the afterglow from this
galaxy, corresponding to about 22 kpc at a redshift of $z$=1.69, does not favor
its identification as the GRB host. Since no galaxy underlying the position of
the afterglow could be detected, only deep flux limits for its host galaxy
could be obtained.  No other host galaxy candidate could be
identified. However, given the redshift of the burst, this is not remarkable
and matches the ensemble properties of the luminosities of GRB host
galaxies found so far \citep{Savaglio2009}. 

\object{GRB 080928} has shown once more the tremendous amount of information that 
can be gathered for a single burst and the fundamental importance 
of both timely responses and the joint analysis of all the available data.
It is the combination of gamma-ray, X-ray, and optical/NIR 
data that once more characterizes the golden age of GRB research.

\begin{acknowledgements}

The authors thank the anonymous referee for a very constructive report.
A. Rossi and S.K. acknowledge support by DFG grant Kl 766/11-3 and AR
additionally from the BLANCEFLOR Boncompagni- Ludovisi, n\'ee Bildt
foundation. S.S., D.A.K. and P.F.  acknowledge support by the Th\"uringer
Landessternwarte Tautenburg, Germany, as well as DFG grant Kl 766/16-1.
T.K. acknowledges support by the DFG cluster of excellence 'Origin and
Structure of the Universe'. S.S. acknowledges further support by a Grant of
Excellence from the Icelandic Research Fund. A. Rossi acknowledges 
Fr\'ed\'eric Daigne, Cristiano Guidorzi, Daniele Pierini, and Sandra Savaglio for helpful
discussions. A. Rau and S.K.  acknowledge Re'em Sari for helpful
remarks. D.A.K. acknowledges A. Zeh for fitting scripts. Part of the
funding  for GROND (both hardware and personnel) was generously
granted from the  Leibniz-Prize to Prof. G. Hasinger (DFG grant HA
1850/28-1). This work made use of data supplied by the UK Swift Science
Data Centre at the  University of Leicester. 

\end{acknowledgements}



\newpage

\appendix

\section{The data set}


\begin{table}[hb]
\begin{minipage}[t]{\columnwidth}
\renewcommand{\footnoterule}{}
\centering
\caption{Log of the ROTSE-IIIa telescope observations.}
\begin{tabular}{lrrrc}
\toprule
Time             & Time             & $T_{\rm start}$      & $T_{\rm stop}$ &       $ CR$ Magnitude    \\
(days)           & (s)              & (s)              & (s)        &                  \\
\midrule

0.002160         &     186.7	    &	  132.0        &     263.9  &	   $>18.5$		\\
0.004509         &     389.6	    &	  272.8        &     556.3  &	   $18.38\pm0.22$	\\
0.008266         &     714.2	    &	  565.6        &     901.8  &	   $17.35\pm0.10$	\\
0.012341     	 &     1066.3	    &	  911.2        &     1247.7 &	   $17.16\pm0.09$	\\
0.016375     	 &     1414.8	    &	  1256.7       &     1592.9 &	   $17.48\pm0.13$	\\
0.020442     	 &     1766.2	    &	  1602.1       &     1947.2 &	   $17.20\pm0.10$	\\
0.024512     	 &     2117.8	    &	  1956.1       &     2293.0 &	   $16.50\pm0.06$	\\
0.028530     	 &     2465.0	    &	  2302.2       &     2639.3 &	   $16.51\pm0.10$	\\
0.034383     	 &     2970.7	    &	  2648.7       &     3331.8 &	   $16.54\pm0.06$	\\
0.042429     	 &     3665.9	    &	  3340.8       &     4022.6 &	   $16.85\pm0.05$	\\
0.054459     	 &     4705.3	    &	  4373.1       &     5062.8 &	   $17.18\pm0.08$	\\
0.062571     	 &     5406.2	    &	  5071.8       &     5762.6 &	   $17.49\pm0.10$	\\
0.070628     	 &     6102.2	    &	  5771.4       &     6452.0 &	   $17.35\pm0.08$	\\
0.078676     	 &     6797.6	    &	  6461.6       &     7151.1 &	   $16.99\pm0.09$	\\
0.087760     	 &     7582.5	    &	  7160.1       &     8029.8 &	   $16.90\pm0.07$	\\
0.096907     	 &     8372.8	    &	  8038.5       &     8720.9 &	   $16.96\pm0.08$	\\
0.104911     	 &     9064.3	    &	  8729.8       &     9411.6 &	   $16.85\pm0.05$	\\
0.112957     	 &     9759.5	    &	  9420.7       &     10110.5&	   $16.68\pm0.06$	\\
0.121012     	 &     10455.4      &	  10119.7      &     10802.3&	   $16.86\pm0.05$	\\
0.130107     	 &     11241.2      &	  10811.6      &     11687.9&	   $16.97\pm0.06$	\\
0.139266     	 &     12032.6      &	  11696.8      &     12378.0&	   $17.17\pm0.07$	\\
0.147299     	 &     12726.6      &	  12386.8      &     13075.8&	   $17.22\pm0.10$	\\
0.155384     	 &     13425.2      &	  13084.4      &     13774.8&	   $17.38\pm0.15$	\\

\bottomrule
\end{tabular}
\tablefoot{
Magnitudes are Vega magnitudes
(unfiltered $R$-equivalent data, see \citealt{Quimby2006a}), 
not corrected for Galactic extinction
(Sect.~\ref{sec:optnir}). 
Midtimes have been derived logarithmically.}
\label{tab:magsROTSE}
\end{minipage}
\end{table}


\newpage

\begin{table}
\begin{minipage}[t]{\columnwidth}
\renewcommand{\footnoterule}{}
\centering
\caption{Log of the \swift/UVOT observations.}
\renewcommand{\tabcolsep}{4.5pt}
\begin{tabular}{lrrrcr}
\toprule
Time         & Time & $T_{\rm start}$ & $T_{\rm stop}$ & Magnitude  & Filter  \\
(days)       & (s)  &  (s)        &       (s)  &            &      \\
\midrule
0.001909     	 &  	164.9   	 & 	160.4   	  &	 169.7    &	 $>$ 17.1	  &	 $v$    \\
0.002184     	 &  	188.7   	 & 	179.0   	  &	 199.0    &  $>$ 19.7	  &	 $white$    \\
0.002416     	 &  	208.7   	 & 	199.0   	  &	 219.0    &	 $19.03\pm0.25$	  &	 $white$    \\
0.002648     	 &  	228.7   	 & 	219.0   	  &	 239.0    &	 $18.70\pm0.21$	  &	 $white$    \\
0.002879     	 &  	248.8   	 & 	239.0   	  &	 259.0    &	 $18.44\pm0.17$	  &	 $white$    \\
0.003110     	 &  	268.7   	 & 	259.0   	  &	 278.7    &	 $19.29\pm0.31$	  &	 $white$    \\
0.003834     	 &  	331.2   	 & 	285.0   	  &	 385.0    &	 $19.15\pm0.41$	  &	 $v$    \\
0.004969     	 &  	429.3   	 & 	385.0   	  &	 478.7    &	 $18.96\pm0.39$	  &	 $v$    \\
0.050063     	 &  	4325.4   	 & 	4226.7  	  &	 4426.5   &	 $>$ 19.1         &	 $uvm2$    \\
0.052440     	 &  	4530.8   	 & 	4432.0  	  &	 4631.8   &	 $18.02\pm0.15$   &      $uvw1$    \\
0.054817     	 &  	4736.2   	 & 	4637.3  	  &	 4837.1   &	 $17.54\pm0.07$   &      $u$    \\
0.057191     	 &  	4941.3   	 & 	4842.5  	  &	 5042.2   &	 $18.05\pm0.08$   &      $b$    \\
0.059573     	 &  	5147.1   	 & 	5048.2  	  &	 5247.9   &	 $17.96\pm0.04$   &      $white$    \\
0.061956     	 &  	5353.0   	 & 	5254.0  	  &	 5453.8   &	 $>$ 19.3         &	 $uvw2$    \\
0.064332     	 &  	5558.3   	 & 	5459.3  	  &	 5659.1   &	 $17.66\pm0.10$	  &	 $v$    \\
0.066706     	 &  	5763.4   	 & 	5664.3  	  &	 5864.1   &	 $>$ 19.0         &	 $uvm2$    \\
0.069081     	 &  	5968.6   	 & 	5869.6  	  &	 6069.3   &	 $18.07\pm0.15$   &      $uvw1$    \\
0.071451     	 &  	6173.4   	 & 	6074.3  	  &	 6274.1   &	 $17.74\pm0.11$   &      $u$    \\
0.073834     	 &  	6379.2   	 & 	6280.1  	  &	 6479.9   &	 $18.02\pm0.08$   &	 $b$    \\
0.076205     	 &  	6584.1   	 & 	6485.0  	  &	 6684.7   &	 $17.81\pm0.05$   &	 $white$    \\
0.078267     	 &  	6762.3   	 & 	6690.7  	  &	 6834.6   &	 $>$ 18.8         &	 $uvw2$    \\
0.117545     	 &  	10155.9   	 & 	10007.1 	  &	 10306.9  &	 $17.16\pm0.07$   &	 $v$    \\
0.121060     	 &  	10459.6   	 & 	10310.8 	  &	 10610.6  &	 $17.24\pm0.06$   &	 $v$    \\
0.124573     	 &  	10763.1   	 & 	10614.3 	  &	 10914.0  &	 $17.34\pm0.07$   &	 $v$    \\
0.131495     	 &  	11361.1   	 & 	10920.2 	  &	 11819.9  &	 $19.66\pm0.28$   &	 $uvm2$    \\
0.141377     	 &  	12214.9   	 & 	11826.8 	  &	 12615.8  &	 $18.28\pm0.10$   &	 $uvw1$    \\
0.184457     	 &  	15937.1   	 & 	15787.9 	  &	 16087.7  &	 $18.35\pm0.09$   &	 $u$    \\
0.187971     	 &  	16240.7   	 & 	16091.5 	  &	 16391.3  &	 $18.21\pm0.09$   &	 $u$    \\
0.191485     	 &  	16544.3   	 & 	16395.1 	  &	 16694.9  &	 $18.38\pm0.09$   &	 $u$    \\
0.195021     	 &  	16849.8   	 & 	16700.6 	  &	 17000.3  &	 $18.74\pm0.08$   &	 $b$    \\
0.198535     	 &  	17153.4   	 & 	17004.2 	  &	 17304.0  &	 $18.87\pm0.09$   &	 $b$    \\
0.202048     	 &  	17456.9   	 & 	17307.7 	  &	 17607.5  &	 $18.71\pm0.09$   &	 $b$    \\
0.205580     	 &  	17762.1   	 & 	17612.9 	  &	 17912.6  &	 $18.51\pm0.06$   &	 $white$    \\
0.209096     	 &  	18065.9   	 & 	17916.6 	  &	 18216.4  &	 $18.64\pm0.06$   &	 $white$    \\
0.211904     	 &  	18308.5   	 & 	18220.2 	  &	 18397.2  &	 $18.65\pm0.09$   &	 $white$    \\
0.254794     	 &  	22014.2   	 & 	21568.9 	  &	 22468.7  &	 $19.25\pm0.13$   &	 $uvw1$    \\
0.261860     	 &  	22624.7   	 & 	22475.3 	  &	 22775.1  &	 $18.45\pm0.10$   &	 $u$    \\
0.265374     	 &  	22928.4   	 & 	22779.0 	  &	 23078.7  &	 $18.75\pm0.12$   &	 $u$    \\
0.268888     	 &  	23231.9   	 & 	23082.5 	  &	 23382.3  &	 $18.74\pm0.13$   &	 $u$    \\
0.272424     	 &  	23537.4   	 & 	23388.0 	  &	 23687.8  &	 $19.25\pm0.14$   &	 $b$    \\
0.275939     	 &  	23841.1   	 & 	23691.7 	  &	 23991.5  &	 $19.11\pm0.16$   &	 $b$    \\
0.278780     	 &  	24086.6   	 & 	23995.4 	  &	 24178.3  &	 $19.49\pm0.30$   &	 $b$    \\
0.321708     	 &  	27795.5   	 & 	27349.3 	  &	 28249.1  &	 $>$ 20.2         &	 $uvm2$    \\
0.332203     	 &  	28702.3   	 & 	28256.0 	  &	 29155.7  &	 $19.71\pm0.18$   &	 $uvw1$    \\
0.339259     	 &  	29312.0   	 & 	29162.5 	  &	 29462.3  &	 $19.34\pm0.23$   &	 $u$    \\
0.342773     	 &  	29615.6   	 & 	29466.1 	  &	 29765.8  &	 $19.73\pm0.37$   &	 $u$    \\
0.345653     	 &  	29864.5   	 & 	29770.2 	  &	 29959.0  &	 $>$ 18.9         &	$u$    \\
0.385349     	 &  	33294.2   	 & 	33157.3 	  &	 33431.6  &	 $>$ 19.6         &	$uvw2$    \\
0.841783     	 &  	72730.0   	 & 	46011.4 	  &	 114964.1 &	 $>$ 21.9         &	$uvw2$    \\
0.853404     	 &  	73734.1   	 & 	46917.8 	  &	 115877.5 &	 $21.14\pm0.31$   &	 $v$    \\
0.903886     	 &  	78095.7   	 & 	52273.5 	  &	 116673.6 &	 $>$ 21.5         &      $uvm2$    \\
0.954216     	 &  	82444.2   	 & 	56280.7 	  &	 120770.5 &	 $>$ 21.8         &      $uvw1$    \\
1.005800     	 &  	86901.3   	 & 	62061.1 	  &	 121684.1 &	 $21.31\pm0.22$	  &	 $u$    \\
1.095660     	 &  	94665.0   	 & 	73622.3 	  &	 121722.0 &	 $>$ 21.3         &	 $b$    \\
0.864812     	 &  	74719.8   	 & 	74534.4 	  &	 74905.6  &	 $21.85\pm0.36$   &      $white$    \\
0.864400     	 &  	74684.2   	 & 	74534.4 	  &	 74834.2  &	 $21.73\pm0.34$   &      $white$    \\
1.811460     	 &  	156510.0   	 & 	121764.2	  & 	201170.5  & 	$>$ 20.5          &      $v$    \\
2.194020     	 &  	189563.0   	 & 	178401.4	  & 	201423.3  & 	$>$ 21.1          &      $u$    \\
2.196340     	 &  	189764.0   	 & 	178556.9	  & 	201674.8  & 	$>$ 21.4          &      $b$    \\
2.199990     	 &  	190079.0   	 & 	178715.3	  & 	202165.7  & 	$>$ 21.3          &      $uvw1$    \\
3.177100     	 &  	274501.0   	 & 	260418.7	  & 	289345.7  & 	$>$ 21.9          &      $u$    \\

\bottomrule
\end{tabular}
\tablefoot{Magnitudes are Vega magnitudes, not corrected for Galactic extinction
(Sect.~\ref{sec:optnir}). Midtimes have been derived logarithmically.}
\label{tab:magsUVOT}
\end{minipage}
\end{table}

\newpage


\begin{table*}
\begin{minipage}[t]{\textwidth}
\renewcommand{\footnoterule}{}
\centering
\caption{Log of the GROND multi-color observations.}
\begin{tabular}{l|c|c|c}
\toprule
Time      & Filter & Exposure & Brightness (mag$_{\rm AB}$) \\
(days)    &        & (s)      &              \\
\midrule
 0.6031  & $g^\prime r^\prime i^\prime z^\prime $ & $12\times370$ & $21.40\pm0.15$ / $21.03\pm0.07$ /	$20.54\pm0.07$ / $20.43\pm0.08$ \\
 0.6031  & $JHK_S$    & $240\times10$ & $19.83\pm0.10$ / $19.49\pm0.15$ /  $19.11\pm0.15$                  \\[1mm]  
 0.7398  & $g^\prime r^\prime i^\prime z^\prime $ & $12\times370$ & $21.93\pm0.16$ / $21.48\pm0.08$ /	$21.13\pm0.10$ / $20.78\pm0.10$ \\
 0.7398  & $JHK_S$    & $360\times10$ & $20.40\pm0.16$ / $20.01\pm0.22$ /  $19.60\pm0.30$                  \\ [1mm]   
 1.7370  & $g^\prime r^\prime i^\prime z^\prime $ & $12\times370$ & $23.35\pm0.25$ / $22.99\pm0.10$ /	$22.56\pm0.16$ / $22.53\pm0.16$ \\
 1.7370  & $JHK_S$    & $360\times10$ & $>21.8$	      / $>20.9$	       /  $>20.4$                         \\[1mm]  
 2.708   & $g^\prime r^\prime i^\prime z^\prime $ & $4\times370$  & $>24.3$	      / $23.41\pm0.40$ /	$23.35\pm0.73$ / $>23.2$	      \\
 2.708	 & $JHK_S$    & $120\times10$ & $>21.2$  	    / $>20.4$  	     /  $>19.8$	                        \\[1mm]  
 201     & $g^\prime r^\prime i^\prime z^\prime $ & $12\times370$ & $>25.4$	      / $>25.6$        /	$>24.6$        / $>24.3$	      \\
 201	 & $JHK_s$    & $360\times10$ & $>22.0$  	    / $>21.6$  	     /  $>20.9$	                        \\ 
\bottomrule
\end{tabular}
\tablefoot{Magnitudes are given in the AB photometric system, not corrected
for Galactic extinction (Sect.~\ref{sec:optnir}). Midtimes have been derived logarithmically.}

\label{tab:mags}
\end{minipage}
\end{table*}


\begin{table*}[htb]
\begin{minipage}[t]{\textwidth}
\renewcommand{\footnoterule}{}
\centering
\caption{Secondary standard stars within 4 arcmin of the afterglow position
(Fig.~\ref{fig:finding}).}
\begin{tabular}{r|c|ccccccc}
\toprule 
\# & R.A., Dec. (J2000) &  $g^\prime $  &  $r^\prime $  &  $i^\prime $ &  $z^\prime $  &  $J$  & $H$  &  $K_s$ \\
\midrule
 1 & 06:20:15.23 $-$55:12:45.4  & 14.426(01) & 13.727(01) & 13.269(01) & 13.102(01) & 12.753(01) & 12.613(02) & 12.880(02)\\
 2 & 06:20:13.45 $-$55:12:32.5  & 19.427(05) & 18.987(05) & 18.593(05) & 18.478(08) & 18.267(05) & 18.219(10) & 18.435(23)\\
 3 & 06:20:13.87 $-$55:12:17.1  & 17.291(01) & 16.982(01) & 16.709(01) & 16.668(02) & 16.568(03) & 16.673(03) & 16.780(07)\\
 4 & 06:20:14.65 $-$55:12:01.1  & 17.513(02) & 17.366(02) & 17.103(02) & 17.072(03) & 16.988(03) & 17.100(03) & 17.208(10)\\
 5 & 06:20:12.70 $-$55:11:55.1  & 20.734(14) & 19.460(08) & 18.017(03) & 17.450(04) & 16.886(03) & 16.778(04) & 16.853(07)\\
 6 & 06:20:12.21 $-$55:11:45.9  & 18.307(02) & 18.061(03) & 17.734(03) & 17.643(04) & 17.508(04) & 17.509(05) & 17.711(23)\\
 7 & 06:20:14.51 $-$55:11:45.1  & 19.989(08) & 18.962(05) & 18.204(04) & 17.946(05) & 17.561(03) & 17.341(05) & 17.378(11)\\
 8 & 06:20:06.18 $-$55:12:02.1  & 20.499(05) & 19.456(03) & 18.453(03) & 18.077(02) & 17.598(06) & 17.353(08) & 17.556(05)\\
 9 & 06:19:58.96 $-$55:12:57.4  & 17.430(03) & 17.413(02) & 17.206(02) & 17.185(01) & 17.123(04) & 17.312(07) & 17.553(07)\\
 10& 06:19:58.75 $-$55:10:40.3  & 19.275(03) & 18.121(02) & 17.026(02) & 16.613(01) & 16.110(05) & 16.022(05) & 16.143(02)\\
 11& 06:19:56.64 $-$55:09:57.4  & 20.949(14) & 20.212(14) & 19.672(13) & 19.389(14) & 18.794(08) & 18.609(12) & 18.403(13)\\
 12& 06:20:16.00 $-$55:10:28.9  & 18.087(03) & 17.755(02) & 17.442(03) & 17.370(01) & 17.168(05) & 17.218(06) & 17.313(04)\\
\bottomrule
\end{tabular}
\tablefoot{ Numbers in parentheses give the
photometric $1\sigma$ statistical uncertainty of the secondary standards 
in units of 10 milli-mag.}
\label{tab:std}
\end{minipage}
\end{table*}

\newpage



\begin{figure*}
\includegraphics[width=18.4cm]{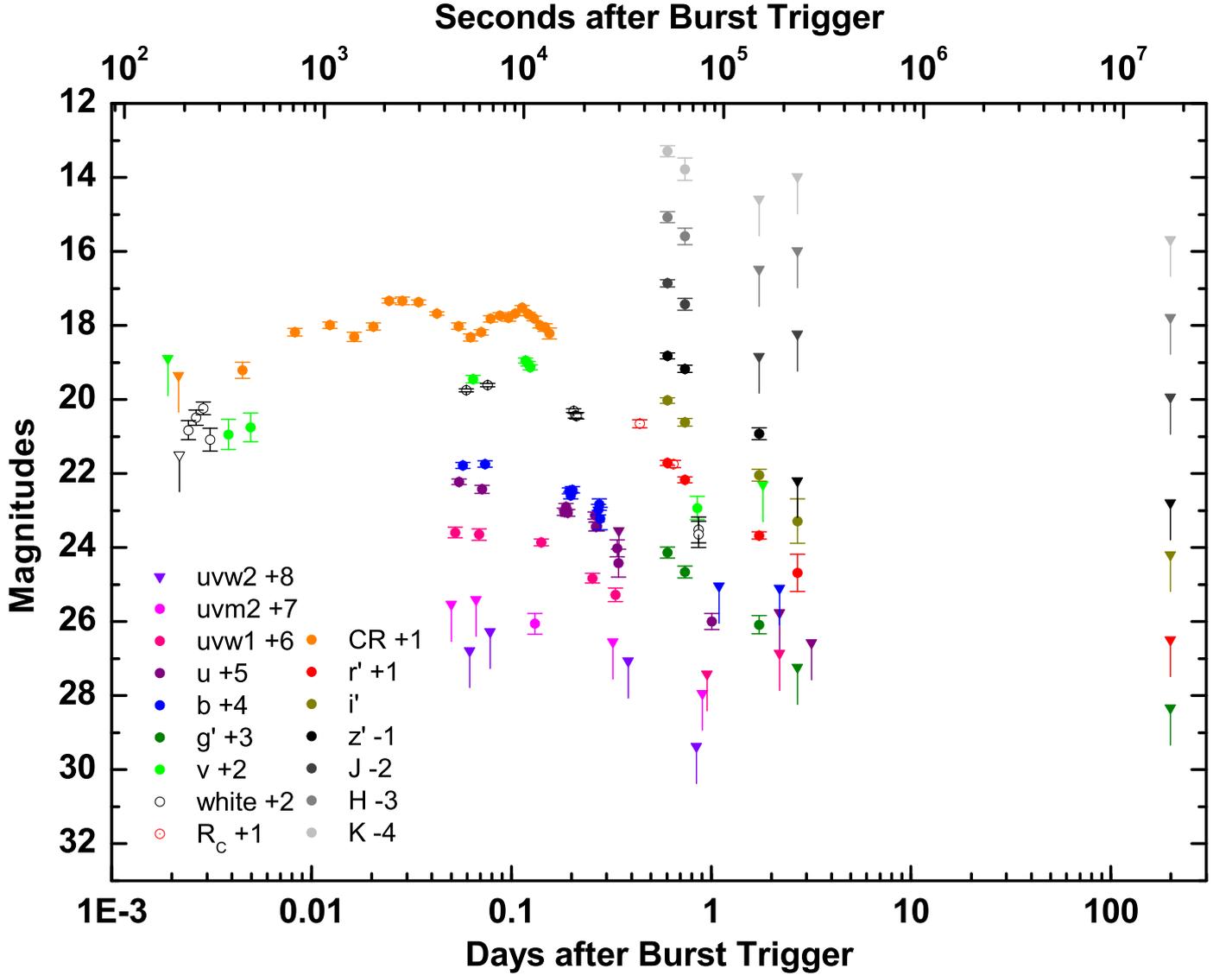}
\caption{The complete optical/NIR data set  of the afterglow of \object{GRB 080928} as
  listed in  Tables~\ref{tab:magsROTSE},~\ref{tab:magsUVOT}, and
  \ref{tab:mags}. All magnitudes are given in the Vega system, and the GROND magnitudes are 
  corrected according \citet{Greiner2008a}. Colors have been shifted by the values given in the legend
  for clarity. Downward pointing triangles are upper limits, $uvw2$ was the
  only filter in which only upper limits could be derived.}
\label{fig:lc_multicolorAlex}
\end{figure*}


\end{document}